# The expensive son hypothesis

Lucas Invernizzi[1*], Jean-François Lemaître[1], Mathieu Douhard[1]

**Institution** : [1]Université de Lyon, Université Lyon 1, CNRS, Laboratoire de Biométrie et Biologie Evolutive UMR 5558, F-69622 Villeurbanne, France

***Correspondence**: lucas.invernizzi@univ-lyon1.fr or lucasinvernizzi@yahoo.com

**Conflicts of interest statement**: We declare no conflicts of interest

**Author contributions statement**: Lucas Invernizzi, Jean-François Lemaître and Mathieu Douhard conceived the ideas and designed methodology. Lucas Invernizzi performed the literature research, notably for tables. All authors led the writing of the manuscript, contributed critically to the drafts and gave final approval for publication.

**Data availability statement**: No data have been used in this paper

**Statement on inclusion**: Our study consisted of a detailed and varied literature research to illustrate a concept. No local data collection was therefore carried out and no data were used in this study. Consequently, this study did not involve indigenous researchers from any particular region. However, we plan to take these reservations into account in our future research projects, especially those involving the use of data of a specific region.

**Abstract**

1. In its initial form, the expensive son hypothesis postulates that sons from male-biased sexually dimorphic species require more food during growth than daughters, which ultimately incur fitness costs for mothers predominantly producing and rearing sons.
2. We first dissect the evolutionary framework in which the expensive son hypothesis is rooted, and we provide a critical reappraisal of its differences from other evolutionary theories proposed in the field of sex allocation. Then, we synthesize the current (and absence of) support for the costs of producing and rearing sons on maternal fitness components (future reproduction and survival).
3. Regarding the consequences in terms of future reproduction, we highlight that species with pronounced sexual size dimorphism display a higher cost of sons than of daughters on subsequent reproductive performance, at least in mammals. However, in most studies, the relative fitness costs of producing and rearing sons and daughters can be due to sex-biased maternal allocation strategies rather than differences in energetic demands of offspring, which constitutes an alternative mechanism to the expensive son hypothesis *stricto sensu*.
4. We observe that empirical studies investigating the differential costs of sons and daughters on maternal survival in non-human animals remain rare, especially for long-term survival. Indeed, most studies have investigated the influence of offspring sex (or litter sex ratio) at year T on survival at year T+1, and they rarely provide a support to the expensive son hypothesis. On the contrary, in humans, most studies have focused on the relationship between proportion of sons and maternal lifespan, but these results are inconsistent.
5. Our study highlights new avenues for future research that should provide a comprehensive view of the expensive son hypothesis, by notably disentangling the effects of offspring behaviour from the effect of sex-specific maternal allocation. Moreover, we emphasize that future studies should also embrace the mechanistic side of the expensive son hypothesis, largely neglected so far, by deciphering the physiological pathways linking son's production to maternal health and fitness.

**Keywords:** energy requirements, fitness costs, maternal investment, offspring sex, sexual size dimorphism

## 1 | An overview of the expensive son hypothesis

In the vast majority of species, sex differences in metabolism and growth appear from the early stages of development and persist during the provisioning of young after birth or hatching (Alur, 2019; Fairbairn et al., 2007), resulting in a sexual size dimorphism at the adult stage. While invertebrates and poikilothermic vertebrates have largely female-biased sexual size dimorphism (SSD), mammals and birds often express male-biased SSD overall (Lindenfors et al., 2007). The direction and magnitude of the SSD is driven by the strength of both natural and sexual selection acting on each sex (Darwin, 1871; Lindenfors et al., 2007). Because early growth is likely to have the greatest influence on male reproductive success in polygynous species showing high SSD (Trivers, 1985), sex differences in the energetic costs of producing and rearing sons and daughters should be most frequently found in these species. In his highly influential book on the evolution of parental care, Tim Clutton-Brock (1991) concluded that "*in polygynous birds and mammals, where males are the larger sex, sons grow faster than daughters, require more food, and depress their parents subsequent breeding success or survival to a greater extent*" (p. 227), which we refer to as the expensive son hypothesis. However, while the differences in the energetics costs of producing and rearing sons and daughters were already well documented in the early 1990's (Clutton-Brock, 1991), the fitness consequences were rarely examined at this time.

Here, we first decipher the evolutionary framework within which the expensive son hypothesis is grounded. Then, we provide a critical reappraisal of the studies quantifying the differential fitness costs of producing and rearing sons and daughters. Finally, we identify open questions that may stimulate future research on the links between offspring sex, health and fitness.

**Figure 1: The expensive son hypothesis.** The greater extraction of maternal resources by sons - in species in which the sexual size dimorphism is biased towards males - is responsible for an increased energetic cost for mothers producing and rearing sons. These energetic costs

can result in fitness consequences in terms of both survival and subsequent reproductive success, particularly in poor environmental conditions. The genetic and physiological mechanisms mediating the long-term fitness consequences of mothers producing and rearing sons are yet to be deciphered. Illustrations have been generated on the site https://app.recraft.ai/projects.

## 2 | Back to the evolutionary theories

The expensive son hypothesis is grounded in the life history theory framework serving to explain the costs of reproduction (Clutton-Brock, 1991), and relies on the principle of energy allocation (Cody, 1966). According to this principle, once resources are limited, energy allocation to one biological function (e.g. growth, reproduction, survival insurance mechanisms) curtails the energy allocation to another function, which ultimately leads to resource-based allocation trade-off. A well-known trade-off occurs between current and future reproduction (Stearns, 1992). If parents increase allocation to current reproduction, this can impair their ability to reproduce in the future, as well their survival (e.g. Rivalan et al., 2005). Conversely, if they do not allocate enough energy in current reproduction, the offspring can have poor survival and/or reproductive success. Overall, costs of reproduction are associated with each breeding stage (e.g. gestation and lactation in mammals; egg production, incubation and chick-rearing in birds). Because lactation is the most energetically costly component of female reproduction in many mammals (Clutton-Brock et al., 1989), the costs of rearing offspring should be higher than those producing them. The costs of reproduction can manifest in the short-term, through e.g. a reduction in current reproductive success following a successful reproductive event the previous year (Hamel et al., 2010), or once a certain amount of physiological damages has accumulated over several reproductive events (Kroeger et al., 2018). These cumulative long-term costs of reproduction are often examined in the context of the disposable soma theory (Kirkwood, 1977; Kirkwood & Rose, 1991), which posits that individuals allocating a substantial quantity of resources toward reproduction early in life reduce their allocation to somatic maintenance, which will ultimately promote a reduced lifespan later in life and/or a faster senescence (see Lemaître et al., 2015 for a review in wild populations of vertebrates).

The initial prediction of the expensive son hypothesis is that producing or rearing sons is associated with short-term detrimental consequences on maternal reproduction and survival (Clutton-Brock, 1991). Whether mothers producing and rearing more sons over several reproductive events suffer from a reduced lifespan or impaired reproductive performance in the long-run remains a major area for exploration. An early study of the association between the lifetime number of sons and maternal longevity was conducted by Helle (2002) in a pre-industrial human population. This marked the beginning of a new line of inquiry into the differential costs of sons and daughters over multiple reproductive events. More recently, Douhard et al. (2019, 2020) examined these issues in non-human long-lived mammals.

The expensive son hypothesis posits that differences in the costs of producing and rearing sons and daughters result from sex differences in offspring behaviour (Clutton-Brock, 1991) such as a higher propensity for sons to extract milk from their mother compared to daughters (that may even lead to acute mother-offspring conflicts over food provisioning, see Haig, 2010). Alternatively, mothers may allocate more energy per son than per daughter during gestation and/or lactation and pay the costs in term of fitness, later. Most research on sex allocation in polygynous vertebrates has focused on the Trivers-Willard hypothesis, which proposes that mothers in good condition invest more in the production of sons while those in bad condition invest more in daughters (Trivers & Willard, 1973 see Gosling, 1986 for an example). Trivers (1972) defined parental investment as any parental behaviour that increases the offspring's chance of survival while decreasing the future survival or reproductive success of the parent. A complicating factor here is that the assumptions of Trivers-Willard hypothesis are more likely to hold in highly sexually dimorphic species (Hewison & Gaillard, 1999). However, while under the Trivers-Willard hypothesis the higher cost of sons should be particularly high for mothers in good condition, we may expect to observe the opposite pattern under the expensive son hypothesis (a higher cost of sons for mothers in poor condition) (Clutton-Brock, 1991). On the other hand, Myers (1978) predicted that parents with low resources (either because they are in poor condition or reproduce during poor environmental conditions) should overproduce the cheapest sex to buffer costs in terms of future reproductive success. Currently, it remains unclear whether environmental conditions and maternal condition commonly influence sex differences in the costs of producing and rearing young.

**3 | Relative costs of sons and daughters in terms of future reproduction**

Most mammalian species with pronounced SSD, such as red deer (*Cervus elaphus),* bighorn sheep (*Ovis canadensis),* fallow deer (*Dama dama),* Eastern gray kangaroo (*Macropus giganteus),* African elephant (*Loxodonta africana)* and long-finned pilot whale (*Globicephala melas),* display higher costs of rearing sons than daughters on subsequent reproductive performance (Table 1). On the other hand, the species for which no higher cost of sons has been reported (e.g. Cuvier's gazelle (*Gazella cuvieri*), white rhinoceros (*Ceratotherium simum)* and Columbian ground squirrel (*Urocitellus columbianus*) are among the least dimorphic mammals (Table 1). One unexpected report comes from the Southern elephant seals (*Mirounga leonina*) and its northern relative (*Mirounga angustirostris).* Elephant seals are some of the most sexually dimorphic mammals with adult males being up to four times heavier than females. In addition, elephant seals are extreme capital breeders whose lactation results in extreme loss in female body mass (35% on average over the 24-days lactation period, see Desprez et al., 2018). In capital-breeding mammals, a mother that had produced or weaned a male may require more time to rebuild her energetic reserve and therefore, might be more likely to skip the following reproductive occasion in annual reproducers or delayed the next reproduction in multi-year lactation species (e.g. *Loxodonta Africana* in Table 1). Yet, this was not observed in empirical studies of elephant seals (Le Bœuf et al., 1989; Wilkinson & Van Aarde, 2001). One possible explanation is that the difference in SSD at the end of the lactation period in these species is modest compared with the pronounced difference in adulthood. For example, in the northern elephant seal, males are only 4% heavier than females at weaning (Salogni et al., 2019). Although SSD has been evaluated primarily in adults (Table 1 and 2), it is important to consider SSD at birth and weaning when examining the relative cost of producing and rearing sons and daughters. Post-weaning SSD can be important only for species with prolonged mother-offspring associations, such as mountain goats (*Oreamnos americanus)*. In this species, mothers of 1-year-old males are more likely to take a reproductive pause the following year while newborn sex did not influence subsequent reproduction in the same population (Table 1). This cost of sons could be due to the fact that 1-year-old males are about 12% heavier than females while there was no sexual dimorphism in newborns mass (Festa-Bianchet & Côté, 2008).

Quite strikingly, the patterns observed in macaques (genus *Macaca)* are not aligned with the predictions of the expensive son hypothesis, as birth intervals after the birth of daughters were longer than after the birth of a son, or the probability of giving birth after a daughter in the following breeding season is lower than after a son (Table 1). Furthermore, the repeatability of these results between populations but also within them (e.g. well studied populations of Madingley and Primate Research Center in California, Table 1) make these results robust. Sexual size dimorphism is biased toward larger males in Macaques (Plavcan, 2001), and a likely explanation for the higher cost of daughters is local resource competition between females. Where daughters remain in their natal group and sons disperse, the continued presence of daughters can increase competition for food within the group (Clark, 1978). Females can minimize competition for themselves and their daughters by reducing the recruitment of unrelated immature females into their group through aggression (Silk, 1983). High levels of aggression directed at female infants cause daughters to spend more time in contact with their mother and the subsequent greater frequency of nipple stimulation may inhibit ovulation and delays the next conception (Gomendio, 1990). In contrast, in Killer whales (*Orcinus orca*), where individuals of both sexes may remain in their natal group, the presence of adult sons has a stronger negative influence on the reproductive output of the mother (Weiss et al., 2023). Due to their larger body size, adult males have energetic requirements exceeding those of adult females (Noren, 2011) and this may limit the quantity of prey that adult males can afford to share with group members (Wright et al., 2016). The examples outlined above thus emphasize that the sex differences in the costs of sons and daughters are not necessarily restricted to gestation and lactation periods.

Females facing persistent costs of sons born during previous reproductive events can transfer some of those costs to their current offspring, which further extend the fitness consequences. For instance, in several human populations, the individuals born after an elder brother have lower birth weight than those born after an elder sister (Table 1). This can lead to short- and long-term consequences as birth weight is associated with early survival and health in adulthood (Barker et al., 1989). There is also evidence in animals that the costs of sons can be incurred by the next offspring. For instance, in bighorn sheep, after controlling for yearly environmental variation, it has been found that mothers that weaned a lamb decreased their reproductive effort the following year, especially if they had weaned a son (Martin & Festa-Bianchet, 2010). Mothers ensure their own mass gain and only allocate the excess energy to their young. Consequently, lambs born after a male have lower summer mass gain during lactation than lambs born after a female (Martin & Festa-Bianchet, 2010), which could explain why survival to weaning was about 10-12% lower if the mother had weaned a son rather than a daughter the previous year (Berube et al., 1996).

There is some evidence that sex differences in the reproductive costs of producing and rearing offspring depend on environmental conditions and may be observed only when resources are poor. While mother red deer from the Isle of Rum are less likely to give birth the following season or suffer from 11 days longer birth intervals after producing a son (Clutton-Brock et al., 1981; Froy et al., 2016; but see Gomendio et al., 1990), this decrease of the ability to give birth the following season was not observed in a population living in a relatively productive forest (Bonenfant et al., 2003). Similarly, in the bighorn sheep population of Ram Mountain, where there have been records of high population density and in which ewes are 10% lighter than those of the Sheep River population (Festa-Bianchet et al., 2019), reproductive costs of sons are often found (Table 1). Typically, the sex of the offspring weaned the previous year has a greater effect on lamb survival to 1 year especially during peaks of high population density (Berube et al., 1996). On the contrary, no such costs were found in the Sheep River population (Festa-Bianchet, 1989). In a natural herd of horses (*Equus caballus),* mothers that reared a son during poor years (defined according to mean body condition) showed an increase in the interbirth interval compared to those that reared a daughter (Monard et al., 1997). Similarly, an increase of interbirth intervals after rearing a son has been observed in African elephant (*Loxodonta Africana*) during droughts, corresponding to year with low food availability (Lee & Moss, 1986). Hence, it is perhaps not surprising that no effect of offspring sex was seen in captive giraffes (*Giraffa camelopardalis*) or cattle (*Bos taurus*) (Table 1), which are generally maintained on generous feeding regimens.

Using his own work in red deer as an illustration, Clutton-Brock (1991) suggested that maternal social rank can modulate the relative costs of producing and rearing sons and daughters. Under the expensive son hypothesis, the additional costs of producing and rearing sons could be greater for low-ranking mothers because they do not have priority of access to the best feeding sites and because their body condition is lower than that of high-ranking females (Gomendio et al., 1990). Since the early 90's, studies examining the role of mother social rank on the costs of offspring sex have flourished in non-human primates and it is clear that the direction and the magnitude of the effects are variable across species (Table 1). In some populations of chimpanzee (*Pan troglodytes)* and gorilla (genus *Gorilla*), high-ranking mothers have longer interbirth intervals following births of sons than births of daughters. This pattern agrees with a common prediction of the Trivers-Willard hypothesis that females in better condition would be expected to invest more in sons. In several species of macaques, low-ranking mothers have longer interbirth intervals following births of daughters than births of sons likely because the daughters of low-ranking mothers are subject to more intense aggression than the daughters of high-ranking mothers (Silk, 1983, see also above). Thus, the mechanisms responsible for the modulating effects of maternal social rank on the relative costs of rearing sons and daughters appears different from a strict energetic cost inherent to the expensive son hypothesis.

Most of the studies listed in table 1 are indicative rather than conclusive because correlation does not imply causation. There are only three studies performing experimental manipulation of postnatal offspring sex ratio with contrasting results (i.e. Koskela et al., 2009; Rutkowska et al., 2011; Schwanz & Robert, 2016). Schwanz and Robert (2016) conducted a cross-fostering experiment in a monotocous marsupial (Tammar wallabies, *Macropus eugenii*), in which females reared the opposite sex that they produced to disassociate the effects of producing and rearing offspring of each sex. Their results indicate that rearing the opposite sex to the one originally produced reduces the mother's probability to produce an offspring the following year, irrespective of the sex of offspring. Therefore, albeit indirect, results from this study do not support the expensive son hypothesis. On the contrary, in Bank voles (*Myodes glareolus*), a rodent in which females can breastfed a litter while being pregnant, daily energy expenditure was higher in females pregnant with a male-biased litter and breastfeeding only males' litters compared with females pregnant with a female-biased litter and breastfeeding only females' litters (Rutkowska et al., 2011). These differences in energy expenditure were associated with direct consequences on fitness since the size of the daughter from pregnant mothers of male-biased litters and breastfeeding only males' litters was smaller. Unfortunately, in most mammalian species, mothers refuse to care for cross-fostered offspring, rendering experimental manipulation impossible.

Finally, it is worth noticing that the energetic costs associated with a biased production of sons might contribute to the between-mothers variability in reproductive ageing patterns. Indeed, while females from most vertebrate species suffer from a decline in reproductive performance with age (Lemaître et al., 2020), the age at the onset and the rate of reproductive ageing can differ largely among mothers (Lemaître & Gaillard, 2023). A recent study focused on bighorn sheep highlighted a possible effect of the cumulative cost of sons since mothers that have weaned a majority of sons between 2 and 7 years of age experienced a faster reproductive ageing than those having weaned a majority of daughters (Douhard et al., 2020). Unfortunately, it is difficult to know offspring sex or sex ratio at each reproductive event in wild species and, in some cases, there could be too few births per mother. This is likely why no other study has examined these cumulative costs and multiple questions remain opened.

**Table 1**

## 4 | Relative costs of sons and daughters in terms of survival

Empirical evidence for a higher cost of sons than daughters on maternal survival are rare (Table 2), compared to those on reproduction (Table 1). In red deer, the difference in survival probabilities between low-ranking females that have reared sons (0.45) and those that have reared daughters (0.85) is strong (Gomendio et al., 1990). This species, however, stands out as an exception in non-human animals (Table 2). Up to now, most studies in non-human animals have investigated influence of offspring sex ratio (or litter sex ratio) at year T on survival at year T+1. However, the differential costs of sons and daughters may occur on a shorter timescale. If the foetus is too large to fit through the birth canal easily, both the foetus and the mother could impair maternal survival. Thus, we can expect that the production complications related to the production of a male lead to a higher mortality risk in sexually dimorphic species. A study performed on bonnet macaques (*Macaca radiata*) did not reveal an effect of the offspring sex among mothers dying straight after birth (Silk, 1988). However, the SSD at birth in this species, as in many primate species (Ellis et al., 2013), is relatively weak with an average birth weight of 400g for females and 450g for males (Rao et al., 1998). In human, the higher rate of caesarean section in male foetuses (Eogan, 2003) should encourage the quantification of the offspring sex consequences on maternal survival during the perinatal period in preindustrial populations where medical assistance during childbirth was rudimentary. Caesarean sections are also more frequently found in male than female foetus in cows (Barkema et al., 1992), but such species have often been the subject of artificial selection in order to increase their productivity in terms of meat, resulting in an increase in the size of individuals (Grobet et al., 1997). To date, it remains unknown whether wild mammalian females displaying a high SSD at birth can experience severe complications during birth due to the difficulty of knowing causes of death (but see Andres et al., 2013).

We know much more about the long-term cumulative costs of producing and rearing sons and daughters in terms of maternal longevity. The association between the proportion of sons and the post-menopausal longevity have been extensively studied in women (Table 2), in the context of the expensive son hypothesis. In humans, differences in maternal energy intake suggest that sons require 10% and 7% more energy than daughters during pregnancy and lactation, respectively (da Costa et al., 2010; Tamimi, 2003). However, the human studies listed in Table 2 provide mixed results with half (9 out of 17 studies) supporting the expensive son hypothesis. Given that environmental conditions modulate trade-offs between life history traits, with stronger trade-offs when environmental conditions are unfavourable (Cohen et al., 2020), we might expect that studies finding a cost of sons were those performed in preindustrial rather than contemporary populations, because medical care and birth control were absent, in addition to a limited access to resources. However, the current picture is not so straightforward (Table 2). It is possible that the between-individual heterogeneity in in resource availability within the population mask the cost of sons at the population level (see Van Noordwijk & de Jong, 1986 for a similar point in the context of resource acquisition). A potential factor to account for this heterogeneity is the socioeconomic status of individuals since women with lower socio-economic status often face food shortages and they are not able to increase caloric intake or reduce physical activity when pregnant or lactating (Jasienska, 2020). Two studies show the presence of a negative effect of the number of children or sons on maternal longevity in the poorest social class only (Lycett et al., 2000; Van De Putte et al., 2004). Unfortunately, information on socioeconomic status are often crude and more generally not available for most pre-industrial populations (Nenko et al., 2014). Overall, identifying accurate proxies of nutritional status and its influence on the differential costs and daughters remains one of the major fundamental challenges for human studies. Infant mortality, which integrates a wide range of socio-ecological factors such as nutrition, nursing practices, health and infectious epidemics, could be a promising candidate given costs of sons on maternal longevity have been found in women facing high infant mortality in a pre-industrial Canadian population (Invernizzi et al., 2024).

In some human societies, daughters provide more care than sons to their elderly parent (Grigoryeva, 2017), providing an alternative explanation to the expensive son hypothesis for the higher cost of sons on maternal longevity. To decipher the potentially confounding effects of sociocultural factors, a study examined the differential cumulative costs of sons and daughters on maternal longevity in four species of wild ungulates with high male-biased sexual size dimorphisms but simpler social systems than humans. Yet, no effect of the offspring sex on maternal longevity was found, whatever the species (Douhard et al., 2019). Results of this study are therefore at odds with those of humans. This can be explained by the fact that humans have different life history strategies to other species, notably characterized by an extending post-reproductive lifespan. In long-lived species in which selection did not favour the evolution of post-reproductive lifespan (e.g. ungulates) and where females thus reproduce until death, high survival at late ages should be particularly beneficial for Darwinian fitness. In such species, the fitness cost of producing and rearing sons should be more expressed in terms of impaired future reproduction than survival because a female's fitness is more dependent on her survival than on the reproductive success in a given year (see also Hamel et al., 2010).

**Table 2**

## 5 | Toward an integrative expensive son hypothesis

Currently, it is often unclear whether differential costs of sons and daughters results from differences in the behaviour of offspring (as assumed by the expensive son hypothesis, Clutton-Brock 1991) rather than a sex-specific allocation strategies from the mothers (e.g. Hewison et al., 2005). Some studies have attempted to decipher these two (non-mutually exclusive) pathways. Typically, in Iberian red deer (*Cervus elaphus hispanicus*), females can adjust their milk composition according to the sex of their offspring, with a milk 15% fatter and 12% richer in protein for sons than for daughters (Landete-Castillejos et al., 2005). In Rhesus macaques (*Macaca mulatta*), a similar adjustment is made by females with a milk containing 26% more calories for sons than for daughters (Hinde, 2007). A maternal behavioural adjustment has also been found in several species of several taxa, notably birds in which females provide 43% and 33% more food to their sons than to their daughters in Brown songlark (*Cinclorhamphus cruralis*) and Wandering albatross (*Diomedea exulans*) (see Magrath et al., 2007; Weimerskirch et al., 2000, respectively), or in mammals in which females control the amount of milk given to their offspring by interrupting suckling bouts, such as in giraffes in which suckling bouts terminated by females are longer for sons (Gloneková et al., 2020). Conversely, some studies suggest that sex-differences in offspring behaviour are particularly pronounced. In Coypu (*Myocastor coypus*), a polytocous species, 94% of sons suckle on high-yielding teats, while only 41% of daughters suckle on these teats (L. Gosling et al., 1984). In Plains zebra (*Equus burchellii*), while suckling bouts are mostly terminated by mothers, offspring initiated the resumption of the bout with a resumption time shorter for sons, suggesting that both the mother and the offspring operate on the energy allocation toward offspring (Pluháček et al., 2011). To refine our understanding of the expensive son hypothesis, studies that simultaneously explore both sex differences in nursing behaviour and fitness cost of offspring sex (e.g. Clutton-Brock et al., 1981; Trillmich, 1986; White et al., 2007) are mandatory.

Although the fitness costs of offspring sex are increasingly studied (see Table 1 & 2), the underpinning mechanisms are still unclear, possibly because the expensive son hypothesis in its earliest development was agnostic to any physiological pathways. One potential mechanism for the expensive son hypothesis could involve the depletion of energy reserves from the mother due to the greater energy requirements of sons. If the mother reproduces too soon after producing and rearing a son, then her energy reserves may not have had sufficient time to be fully restored. This may result in the production of offspring with weak growth and poor survival. However, a mother who has produced a son may also wait longer before reproducing again, allowing her energy reserves to be restored. This may result in longer birth interval following the birth of a son. Such a mechanism may explain why sons often reduce maternal subsequent reproduction (Table 2). With regard to the long-term costs of sons, such as reduced maternal longevity, the mechanisms remain to be identified. However, inflammageing, corresponding to an increase in inflammatory factors throughout life without any sign of severe infections disease, and oxidative stress, which are known to be linked to the occurrence of late-onset diseases (Ferrucci & Fabbri, 2018; Finkel & Holbrook, 2000), could play a role. Indeed, two markers of inflammageing (C-reactive protein & Interleukin 6), as well the level of oxidative stress, are both positively correlated with the production of sons (Galbarczyk et al., 2021; Merkling et al., 2017; Ziomkiewicz et al., 2016).

## 6 | Future directions for comparative analyses

A challenge for future research is to determine the quantitative evidence for the expensive son hypothesis. Considering all studies, it would extremely useful to carry out a comparative analysis to test whether there is a statistically significant relationship in the predicted direction, namely that higher costs of sons are observed in populations where there is a greater degree of SSD at weaning. In mammals, lactation is considerably more energy-demanding than gestation (Clutton-Brock et al. 1989, Butte & King, 2005). Therefore, considering lactation, through a standardised SSD at weaning, would enable a more detectable assessment of expensive son hypothesis. Unfortunately, the SSD at weaning is not always known for populations (Table 1 & Table 2) and working with a SSD at the scale of the species could be unsuitable given the variations there may be between populations of a same species (e.g. *Ovis canadensis* populations in Table 1). This gap needs to be filled in the future to better assess whether the costs of sons are higher in highly dimorphic species and if they are reversed in species where the SSD is based towards daughters. Unfortunately, despite a good number of species having a SSD biased towards daughters, such as Spotted hyena (*Crocuta Crocuta*) (Ralls, 1976), there are no studies on these species in our tables, which suggests another gap to be filled in the future.

Furthermore, the metrics employed in previous studies are not easy to standardize. Among the studies that have examined the effect of offspring sex on the mother’s future reproduction, two metrics in particular have been widely used: birth intervals and the probability of giving birth according to the sex of the previous offspring (Table 1). Although there are far fewer studies on maternal survival (Table 2), the probability of maternal survival from year T to T+1 as well maternal longevity are the most widely used metrics. To perform a comparative analysis in the future, more studies on these metrics are necessary.

## 7 | Conclusion

Originally, the expensive son hypothesis referred to the reduced subsequent breeding success or survival of parents due to the higher energy demand of producing males, the largest sex in most mammals and birds (Clutton-Brock, 1991). Overall, empirical studies tackling this hypothesis so far have revealed contrasting results, highlighting marked differences between species or between the fitness related-traits considered (Tables 1 and 2). Moreover, the thorough consideration of many additional factors, not embedded in the expensive son hypothesis so far (e.g. maternal social rank, environmental conditions, lactation stage) could help refining the predictions regarding the occurrence and magnitude of the fitness costs associated with the production of sons and daughters. Finally, we stress that future studies aiming to gain a comprehensive view of the expensive son hypothesis should embrace the putative physiological mechanisms linking maternal reproductive expenditure to fitness.

**Box 1: The expensive son hypothesis: some opened questions**

**Relative costs of sons and daughters in terms of future reproduction**

- Does having a majority of sons lead to an earlier onset or a stronger rate of reproductive senescence for mothers compared to those with a majority of daughters?

**Relative costs of sons and daughters in terms of future survival**

- Does the risk of birth complications for the mother increase with a son in highly dimorphic species?
- What is the most appropriate proxy for considering differences in nutritional status during parental care between mothers when studying the costs of sons on survival in humans and other mammals?

**Mechanisms underpinning the expensive son hypothesis**

- What are the exact molecular pathways (e.g. endocrine pathways) underpinning the expensive son hypothesis and do they differ between species?
- Does inflammageing, recently found to increase with son's production (Galbarczyk et al., 2021), could be an important mechanism involved in the costs of sons on maternal longevity in vertebrates?
- Does the oxidative stress, involved in the aetiology of many diseases and found to be higher in women having a majority of sons (Finkel & Holbrook, 2000; Merkling et al., 2017), could be involved in the costs of sons on maternal longevity in vertebrates?

**Future directions for comparative analyses**

- Given lactation is energetically more costly than gestation, especially in mammals (Butte & King, 2005), does the extent of the expensive son hypothesis is stronger when the relative number of sons is considered at weaning rather than at birth?

- Does the population type could play a role in the detection of costs of sons, with the expectation that costs of sons could be more detectable in populations facing harsh environmental conditions?

## References

Alados, C. L., & Escós, J. M. (1994). Variation in the sex ratio of a low dimorphic polygynous species with high levels of maternal reproductive effort: Cuvier's gazelle. *Ethology Ecology & Evolution*, *6*(3), 301–311. https://doi.org/10.1080/08927014.1994.9522983

Alur, P. (2019). Sex Differences in Nutrition, Growth, and Metabolism in Preterm Infants. *Frontiers in Pediatrics*, *7*, 22. https://doi.org/10.3389/fped.2019.00022

Andres, D., Clutton-Brock, T. H., Kruuk, L. E. B., Pemberton, J. M., Stopher, K. V., & Ruckstuhl, K. E. (2013). Sex differences in the consequences of maternal loss in a long-lived mammal, the red deer (Cervus elaphus). *Behavioral Ecology and Sociobiology*, *67*(8), 1249–1258. https://doi.org/10.1007/s00265-013-1552-3

Bădescu, I., Watts, D. P., Katzenberg, M. A., & Sellen, D. W. (2022). Maternal lactational investment is higher for sons in chimpanzees. *Behavioral Ecology and Sociobiology*, *76*(3), 44. https://doi.org/10.1007/s00265-022-03153-1

Bairagi, R., & Langsten, R. L. (1986). Sex preference for children and its implications for fertility in rural Bangladesh. *Studies in Family Planning*, *17*(6), 302. https://doi.org/10.2307/1966907

Barkema, H. W., Schukken, Y. H., Guard, C. L., Brand, A., & Van Der Weyden, G. C. (1992). Cesarean section in dairy cattle: A study of risk factors. *Theriogenology*, *37*(2), 489–506. https://doi.org/10.1016/0093-691X(92)90206-7

Barker, D. J. P., Osmond, C., Winter, P. D., Margetts, B., & Simmonds, S. J. (1989). Weight in infancy and death from ischaemic heart disease. *The Lancet*, *334*(8663), 577–580. https://doi.org/10.1016/S0140-6736(89)90710-1

Barnier, F., Grange, S., Ganswindt, A., Ncube, H., & Duncan, P. (2012). Inter-birth interval in zebras is longer following the birth of male foals than after female foals. *Acta Oecologica*, *42*, 11–15. https://doi.org/10.1016/j.actao.2011.11.007

Barra, T., Viblanc, V. A., Saraux, C., Murie, J. O., & Dobson, F. S. (2021). Parental investment in the Columbian ground squirrel: Empirical tests of sex allocation models. *Ecology*, *102*(11), e03479. https://doi.org/10.1002/ecy.3479

Beise, J., & Voland, E. (2002). Effect of producing sons on maternal longevity in premodern populations. *Science*, *298*(5592), 317–317. https://doi.org/10.1126/science.298.5592.317a

Bercovitch, F. B., Bashaw, M. J., Penny, C. G., & Rieches, R. G. (2004). Maternal investment in captive giraffes. *Journal of Mammalogy*, *85*, 428–431. https://doi.org/10.1644/1383938

Berman, C. M. (1988). Maternal condition and offspring sex ratio in a group of free-ranging rhesus monkeys: An eleven-year study. *The American Naturalist*, *131*(3), 307–328. https://doi.org/10.1086/284792

Berube, C. H., Festa-Bianchet, M., & Jorgenson, J. T. (1996). Reproductive costs of sons and daughters in Rocky Mountain bighorn sheep. *Behavioral Ecology*, *7*(1), 60–68. https://doi.org/10.1093/beheco/7.1.60

Birgersson, B. (1998). Male-biased maternal expenditure and associated costs in fallow deer. *Behavioral Ecology and Sociobiology*, *43*(2), 87–93. https://doi.org/10.1007/s002650050470

Blanchard, R., & Ellis, L. (2001). Birth weight, sexual orientation and the sex of preceding siblings. *Journal of Biosocial Science*, *33*(3), 451–467. https://doi.org/10.1017/S0021932001004515

Boesch, C. (1997). Evidence for dominant wild female chimpanzees investing more in sons. *Animal Behaviour*, *54*(4), 811–815. https://doi.org/10.1006/anbe.1996.0510

Bonenfant, C., Gaillard, J.-M., Loison, A., & Klein, F. (2003). Sex-ratio variation and reproductive costs in relation to density in a forest-dwelling population of red deer (Cervus elaphus). *Behavioral Ecology*, *14*(6), 862–869. https://doi.org/10.1093/beheco/arg077

Butte, N. F., & King, J. C. (2005). Energy requirements during pregnancy and lactation. *Public Health Nutrition*, *8*(7a), Article 7a. https://doi.org/10.1079/PHN2005793

Cameron, E. Z., & Linklater, W. L. (2000). Individual mares bias investment in sons and daughters in relation to their condition. *Animal Behaviour*, *60*(3), 359–367. https://doi.org/10.1006/anbe.2000.1480

Cesarini, D., Lindqvist, E., & Wallace, B. (2007). Maternal longevity and the sex of offspring in pre-industrial Sweden. *Annals of Human Biology*, *34*(5), Article 5. https://doi.org/10.1080/03014460701517215

Cesarini, D., Lindqvist, E., & Wallace, B. (2009). Is there an adverse effect of sons on maternal longevity? *Proceedings of the Royal Society B: Biological Sciences*, *276*(1664), Article 1664. https://doi.org/10.1098/rspb.2009.0051

Charest Castro, K., Leblond, M., & Côté, S. D. (2018). Costs and benefits of post-weaning associations in mountain goats. *Behaviour*, *155*(4), 295–326. https://doi.org/10.1163/1568539X-00003490

Christiansen, S. G. (2014). The impact of children’s sex composition on parents’ mortality. *BMC Public Health*, *14*(1), 989. https://doi.org/10.1186/1471-2458-14-989

Clark, A. B. (1978). Sex ratio and local resource competition in a prosimian primate. *Science*, *201*(4351), 163–165. https://doi.org/10.1126/science.201.4351.163

Clark, M. M., Bone, S., & Galef, B. G. (1990). Evidence of sex-biased postnatal maternal investment by Mongolian gerbils. *Animal Behaviour*, *39*(4), 735–744. https://doi.org/10.1016/S0003-3472(05)80385-9

Clutton-Brock, T. H. (1991). *The evolution of parental care*. Princeton University Press. https://doi.org/10.1515/9780691206981

Clutton-Brock, T. H., Albon, S. D., & Guinness, F. E. (1981). Parental investment in male and female offspring in polygynous mammals. *Nature*, *289*(5797), Article 5797. https://doi.org/10.1038/289487a0

Clutton-Brock, T. H., Albon, S. D., & Guinness, F. E. (1989). Fitness costs of gestation and lactation in wild mammals. *Nature*, *337*(6204), 260–262. https://doi.org/10.1038/337260a0

Cody, M. L. (1966). A general theory of clutch size. *Evolution*, *20*(2), Article 2. https://doi.org/10.1111/j.1558-5646.1966.tb03353.x

Côté, K., Blanchard, R., & Lalumière, M. L. (2003). The influence of birth order on birth weight: Does the sex of preceding siblings matter? *Journal of Biosocial Science*, *35*(3), 455–462. https://doi.org/10.1017/S0021932003004553

Cohen, A. A., Coste, C. F. D., Li, X., Bourg, S., & Pavard, S. (2020). Are trade-offs really the key drivers of ageing and life span? *Functional Ecology*, *34*(1), 153–166. https://doi.org/10.1111/1365-2435.13444

Crockett, C. M., & Rudran, R. (1987). Red howler monkey birth data II: Interannual, habitat, and sex comparisons. *American Journal of Primatology*, *13*(4), 369–384. https://doi.org/10.1002/ajp.1350130403

da Costa, T. H. M., Haisma, H., Wells, J. C. K., Mander, A. P., Whitehead, R. G., & Bluck, L. J. C. (2010). How much human milk do infants consume? Data from 12 countries using a standardized stable isotope methodology. *The Journal of Nutrition*, *140*(12), Article 12. https://doi.org/10.3945/jn.110.123489

Darwin, C. (1871). *The descent of man, and selection in relation to sex* (Vol. 1).

Desprez, M., Gimenez, O., McMahon, C. R., Hindell, M. A., & Harcourt, R. G. (2018). Optimizing lifetime reproductive output: Intermittent breeding as a tactic for females in a long-lived, multiparous mammal. *Journal of Animal Ecology*, *87*(1), 199–211. https://doi.org/10.1111/1365-2656.12775

Dessouky, F. I., & Rakha, A. H. (1961). Studies on the gestation period and post-partum heat of Friesian cattle in Egypt. *The Journal of Agricultural Science*, *57*(3), 325–327. https://doi.org/10.1017/S0021859600049285

Dhillon, J. S., Acharya, R. M., Tiwana, M. S., & Aggarwal, S. C. (1970). Factors affecting the interval between calving and conception in Hariana cattle. *Animal Science*, *12*(1), 81–87. https://doi.org/10.1017/S0003356100028750

Douhard, M., Festa-Bianchet, M., Hamel, S., Nussey, D. H., Côté, S. D., Pemberton, J. M., & Pelletier, F. (2019). Maternal longevity and offspring sex in wild ungulates. *Proceedings of the Royal Society B: Biological Sciences*, *286*(1896), Article 1896. https://doi.org/10.1098/rspb.2018.1968

Douhard, M., Festa-Bianchet, M., & Pelletier, F. (2020). Sons accelerate maternal aging in a wild mammal. *Proceedings of the National Academy of Sciences*, *117*(9), Article 9. https://doi.org/10.1073/pnas.1914654117

Ellis, L., Hershberger, S., Field, E., Wersinger, S., Pellis, S., Geary, D., Palmer, C., Hoyenga, K., Hetsroni, A., & Karadi, K. (2013). *Sex differences: Summarizing more than a century of scientific research* (0 ed.). Psychology Press. https://doi.org/10.4324/9780203838051

Eogan, M. A. (2003). Effect of fetal sex on labour and delivery: Retrospective review. *BMJ*, *326*(7381), Article 7381. https://doi.org/10.1136/bmj.326.7381.137

Fairbairn, D. J., Blanckenhorn, W. U., & Székely, T. (2007). *Sex, size, and gender roles: Evolutionary studies of sexual size dimorphism*. Oxford University Press.

Ferrucci, L., & Fabbri, E. (2018). Inflammageing: Chronic inflammation in ageing, cardiovascular disease, and frailty. *Nature Reviews Cardiology*, *15*(9), 505–522. https://doi.org/10.1038/s41569-018-0064-2

Festa-Bianchet, M. (1989). Individual differences, parasites, and the costs of reproduction for bighorn ewes (Ovis canadensis). *The Journal of Animal Ecology*, *58*(3), 785. https://doi.org/10.2307/5124

Festa-Bianchet, M., & Côté, S. D. (2008). *Mountain goats: Ecology, behavior, and conservation of an alpine ungulate*. Island Press : Made available through hoopla.

Festa-Bianchet, M., Côté, S. D., Hamel, S., & Pelletier, F. (2019). Long-term studies of bighorn sheep and mountain goats reveal fitness costs of reproduction. *Journal of Animal Ecology*, *88*(8), 1118–1133. https://doi.org/10.1111/1365-2656.13002

Finkel, T., & Holbrook, N. J. (2000). Oxidants, oxidative stress and the biology of ageing. *Nature*, *408*(6809), 239–247. https://doi.org/10.1038/35041687

Froy, H., Walling, C. A., Pemberton, J. M., Clutton-Brock, T., & Kruuk, L. E. B. (2016, September). *Relative costs of offspring sex and offspring survival in a polygynous mammal | Biology Letters*. https://royalsocietypublishing.org/doi/full/10.1098/rsbl.2016.0417

Galbarczyk, A., Klimek, M., Blukacz, M., Nenko, I., Jabłońska, M., & Jasienska, G. (2021). Inflammaging: Blame the sons. Relationships between the number of sons and the level of inflammatory mediators among post-reproductive women. *American Journal of Physical Anthropology*, *175*(3), 656–664. https://doi.org/10.1002/ajpa.24295

Gloneková, M., Brandlová, K., & Pluháček, J. (2020). Giraffe males have longer suckling bouts than females. *Journal of Mammalogy*, *101*(2), 558–563. https://doi.org/10.1093/jmammal/gyz201

Gomendio, M. (1990). The influence of maternal rank and infant sex on maternal investment trends in rhesus macaques: Birth sex ratios, inter-birth intervals and suckling patterns. *Behavioral Ecology and Sociobiology*, *27*(5), 365–375. https://doi.org/10.1007/BF00164008

Gomendio, M., Clutton-Brock, T. H., Albon, S. D., Guinness, F. E., & Simpson, M. J. (1990). Mammalian sex ratios and variation in costs of rearing sons and daughters. *Nature*, *343*(6255), Article 6255. https://doi.org/10.1038/343261a0

Gosling, L., Baker, S., & Wright, K. (1984). Differential investment by female Coypus (Myocastor coypus) during lactation. *Symposium of the Zoological Society of London*, *51*, 273–300.

Gosling, L. M. (1986). Selective abortion of entire litters in the coypu: Adaptive control of offspring production in relation to quality and sex. *The American Naturalist*, *127*(6), 772–795. https://doi.org/10.1086/284524

Grandi, S. M., Hinkle, S. N., Mumford, S. L., Sjaarda, L. A., Grantz, K. L., Mendola, P., Mills, J. L., Pollack, A. Z., Yeung, E., Zhang, C., & Schisterman, E. F. (2023). Infant sex at birth and long-term maternal mortality. *Paediatric and Perinatal Epidemiology*, *37*(3), 229–238. https://doi.org/10.1111/ppe.12933

Green, W. C. H., & Rothstein, A. (1991). Sex bias or equal opportunity? Patterns of maternal investment in bison. *Behavioral Ecology and Sociobiology*, *29*(5), 373–384. https://doi.org/10.1007/BF00165963

Grigoryeva, A. (2017). Own gender, sibling's gender, parent's gender: The division of elderly parent care among adult children. *American Sociological Review*, *82*(1), 116–146. https://doi.org/10.1177/0003122416686521

Grobet, L., Royo Martin, L. J., Poncelet, D., Pirottin, D., Brouwers, B., Riquet, J., Schoeberlein, A., Dunner, S., Ménissier, F., Massabanda, J., Fries, R., Hanset, R., & Georges, M. (1997). A deletion in the bovine myostatin gene causes the double–muscled phenotype in cattle. *Nature Genetics*, *17*(1), 71–74. https://doi.org/10.1038/ng0997-71

Haig, D. (2010). Transfers and transitions: Parent–offspring conflict, genomic imprinting, and the evolution of human life history. *Proceedings of the National Academy of Sciences*, *107*(suppl_1), 1731–1735. https://doi.org/10.1073/pnas.0904111106

Hamel, S., Côté, S. D., & Festa-Bianchet, M. (2011). Tradeoff between offspring mass and subsequent reproduction in a highly iteroparous mammal. *Oikos*, *120*(5), 690–695. https://doi.org/10.1111/j.1600-0706.2011.19382.x

Hamel, S., Gaillard, J.-M., Yoccoz, N. G., Loison, A., Bonenfant, C., & Descamps, S. (2010). Fitness costs of reproduction depend on life speed: Empirical evidence from mammalian populations. *Ecology Letters*, *13*(7), Article 7. https://doi.org/10.1111/j.1461-0248.2010.01478.x

Harrell, C. J., Smith, K. R., & Mineau, G. P. (2008). Are girls good and boys bad for parental longevity?: The effects of sex composition of offspring on parental mortality past age 50. *Human Nature*, *19*(1), Article 1. https://doi.org/10.1007/s12110-008-9028-2

Helle, S., & Lummaa, V. (2013). A trade-off between having many sons and shorter maternal post-reproductive survival in pre-industrial Finland. *Biology Letters*, *9*(2), Article 2. https://doi.org/10.1098/rsbl.2013.0034

Helle, S., Lummaa, V., & Jokela, J. (2002). Sons reduced maternal longevity in preindustrial humans. *Science*, *296*(5570), 1085–1085. https://doi.org/10.1126/science.1070106

Helle, S., Lummaa, V., & Jokela, J. (2010). On the number of sons born and shorter lifespan in historical Sami mothers. *Proceedings of the Royal Society B: Biological Sciences*, *277*(1696), Article 1696. https://doi.org/10.1098/rspb.2009.2114

Hewison, A. J. M., Gaillard, J., Kjellander, P., Toïgo, C., Liberg, O., & Delorme, D. (2005). Big mothers invest more in daughters – reversed sex allocation in a weakly polygynous mammal. *Ecology Letters*, *8*(4), 430–437. https://doi.org/10.1111/j.1461-0248.2005.00743.x

Hewison, A. J. M., & Gaillard, J.-M. (1999). Successful sons or advantaged daughters? The Trivers–Willard model and sex-biased maternal investment in ungulates. *Trends in Ecology & Evolution*, *14*(6), 229–234. https://doi.org/10.1016/S0169-5347(99)01592-X

Hinde, K. (2007). First-time macaque mothers bias milk composition in favor of sons. *Current Biology*, *17*(22), R958–R959. https://doi.org/10.1016/j.cub.2007.09.029

Hinde, K. (2009). Richer milk for sons but more milk for daughters: Sex-biased investment during lactation varies with maternal life history in rhesus macaques. *American Journal of Human Biology*, *21*(4), 512–519. https://doi.org/10.1002/ajhb.20917

Hogg, JohnT., Hass, ChristineC., & Jenni, DonaldA. (1992). Sex-biased maternal expenditure in Rocky Mountain bighorn sheep. *Behavioral Ecology and Sociobiology*, *31*(4). https://doi.org/10.1007/BF00171679

Hurt, L. S., Ronsmans, C., & Quigley, M. (2006). Does the number of sons born affect long-term mortality of parents? A cohort study in rural Bangladesh. *Proceedings of the*

*Royal Society B: Biological Sciences*, *273*(1583), Article 1583. https://doi.org/10.1098/rspb.2005.3270

Invernizzi, L., Bergeron, P., Pelletier, F., Lemaître, J.-F., & Douhard, M. (2024). Sons shorten mother's lifespan in pre-industrial families with high level of infant mortality. *The American Naturalist*, 731792. https://doi.org/10.1086/731792

Jasienska, G. (2020). Costs of reproduction and ageing in the human female. *Philosophical Transactions of the Royal Society B: Biological Sciences*, *375*(1811), Article 1811. https://doi.org/10.1098/rstb.2019.0615

Jasienska, G., Nenko, I., & Jasienski, M. (2006). Daughters increase longevity of fathers, but daughters and sons equally reduce longevity of mothers. *American Journal of Human Biology*, *18*(3), Article 3. https://doi.org/10.1002/ajhb.20497

Jelenkovic, A., Silventoinen, K., Tynelius, P., Helle, S., & Rasmussen, F. (2014). Sex of preceding sibling and anthropometrics of subsequent offspring at birth and in young adulthood: A population-based study in Sweden. *American Journal of Physical Anthropology*, *154*(4), 471–478. https://doi.org/10.1002/ajpa.22534

Kanaziz, R., Huyvaert, K. P., Wells, C. P., Van Vuren, D. H., & Aubry, L. M. (2022). Maternal survival costs in an asocial mammal. *Ecology and Evolution*, *12*(5), e8874. https://doi.org/10.1002/ece3.8874

Kim, J. (2022). The relationship between children's gender composition and parents' all-cause mortality among older adults in Korea. *Journal of Applied Gerontology*, *41*(3), 754–762. https://doi.org/10.1177/07334648211012122

Kirkwood, T. B. L. (1977). Evolution of ageing. *Nature*, *270*(5635), 301–304. https://doi.org/10.1038/270301a0

Kirkwood, T. B. L., & Rose, M. R. (1991). Evolution of senescence: Late survival sacrificed for reproduction. *Philosophical Transactions of the Royal Society of London. Series B: Biological Sciences*, *332*(1262), Article 1262. https://doi.org/10.1098/rstb.1991.0028

Korsten, P., Clutton-Brock, T., Pilkington, J. G., Pemberton, J. M., & Kruuk, L. E. B. (2009). Sexual conflict in twins: Male co-twins reduce fitness of female Soay sheep. *Biology Letters*, *5*(5), 663–666. https://doi.org/10.1098/rsbl.2009.0366

Koskela, E., Mappes, T., Niskanen, T., & Rutkowska, J. (2009). Maternal investment in relation to sex ratio and offspring number in a small mammal – a case for Trivers and Willard theory? *Journal of Animal Ecology*, *78*(5), 1007–1014. https://doi.org/10.1111/j.1365-2656.2009.01574.x

Kroeger, S. B., Blumstein, D. T., Armitage, K. B., Reid, J. M., & Martin, J. G. A. (2018). Cumulative reproductive costs on current reproduction in a wild polytocous mammal. *Ecology and Evolution*, *8*(23), Article 23. https://doi.org/10.1002/ece3.4597

Kurita, H., Matsui, T., Shimomura, T., Fujita, T., & Oka, T. (2012). Maternal investment in sons and daughters in provisioned, free-ranging Japanese macaques (*Macaca fuscata*). *Anthropological Science*, *120*(1), 33–38. https://doi.org/10.1537/ase.110120

Landete-Castillejos, T., Garcia, A., Lopez-Serrano, F. R., & Gallego, L. (2005). Maternal quality and differences in milk production and composition for male and female Iberian red deer calves (Cervus elaphus hispanicus). *Behavioral Ecology and Sociobiology*, *57*(3), 267–274. https://doi.org/10.1007/s00265-004-0848-8

Le Bœuf, B. J., Condit, R., & Reiter, J. (1989). Parental investment and the secondary sex ratio in northern elephant seals. *Behavioral Ecology and Sociobiology*, *25*(2), 109–117. https://doi.org/10.1007/BF00302927

Lee, P. C., & Moss, C. J. (1986). Early maternal investment in male and female African elephant calves. *Behavioral Ecology and Sociobiology*, *18*(5), 353–361. https://doi.org/10.1007/BF00299666

Lemaître, J.-F., Berger, V., Bonenfant, C., Douhard, M., Gamelon, M., Plard, F., & Gaillard, J.-M. (2015). Early-late life trade-offs and the evolution of ageing in the wild. *Proceedings of the Royal Society B: Biological Sciences*, *282*(1806), Article 1806. https://doi.org/10.1098/rspb.2015.0209

Lemaître, J.-F., & Gaillard, J.-M. (2023). Editorial: The evolutionary roots of reproductive ageing and reproductive health across the tree of life. *Frontiers in Ecology and Evolution*, *11*, 1349845. https://doi.org/10.3389/fevo.2023.1349845

Lemaître, J.-F., Ronget, V., & Gaillard, J.-M. (2020). Female reproductive senescence across mammals: A high diversity of patterns modulated by life history and mating traits. *Mechanisms of Ageing and Development*, *192*, 111377. https://doi.org/10.1016/j.mad.2020.111377

Lindenfors, P., Gittleman, J. L., & Jones, K. E. (2007). Sexual size dimorphism in mammals. In *Sex, size, and gender roles: Evolutionary studies of sexual size dimorphism* (pp. 16–27). Oxford University Press.

Lummaa, V. (2001). Reproductive investment in pre–industrial humans: The consequences of offspring number, gender and survival. *Proceedings of the Royal Society of London. Series B: Biological Sciences*, *268*(1480), Article 1480. https://doi.org/10.1098/rspb.2001.1786

Lunn, N. J., & Arnould, J. P. Y. (1997). Maternal investment in Antarctic fur seals: Evidence for equality in the sexes? *Behavioral Ecology and Sociobiology*, *40*(6), 351–362. https://doi.org/10.1007/s002650050351

Lycett, J. E., Dunbar, R. I. M., & Voland, E. (2000). Longevity and the costs of reproduction in a historical human population. *Proceedings of the Royal Society of London. Series B: Biological Sciences*, *267*(1438), Article 1438. https://doi.org/10.1098/rspb.2000.0962

Mace, R., & Sear, R. (1997). Birth interval and the sex of children in a traditional African population: An evolutionnary analysis. *Journal of Biosocial Science*, *29*(4), Article 4. https://doi.org/10.1017/S0021932097004999

Maestripieri, D. (2001). Female-biased maternal investment in rhesus macaques. *Folia Primatologica*, *72*(1), 44–47. https://doi.org/10.1159/000049920

Maestripieri, D. (2002). Maternal dominance rank and age affects offspring sex ratio in pigtail macaques. *Journal of Mammalogy*, *83*(2), 563–568. https://doi.org/10.1644/1545-1542(2002)083<0563:MDRAAA>2.0.CO;2

Magnus, P., Berg, K., & Bjérkedal, T. (1985). The association of parity and birth weight: Testing the sensitization hypothesis. *Early Human Development*, *12*(1), 49–54. https://doi.org/10.1016/0378-3782(85)90136-7

Magrath, M. J. L., Van Lieshout, E., Pen, I., Visser, G. H., & Komdeur, J. (2007). Estimating expenditure on male and female offspring in a sexually size-dimorphic bird: A comparison of different methods. *Journal of Animal Ecology*, *76*(6), 1169–1180. https://doi.org/10.1111/j.1365-2656.2007.01292.x
Margulis, SusanW., Altmann, J., & Ober, C. (1993). Sex-biased lactational duration in a human population and its reproductive costs. *Behavioral Ecology and Sociobiology*, *32*(1), Article 1. https://doi.org/10.1007/BF00172221
Martin, J. G. A., & Festa-Bianchet, M. (2010). Bighorn ewes transfer the costs of reproduction to their lambs. *The American Naturalist*, *176*(4), 414–423. https://doi.org/10.1086/656267
McArdle, P. F., Pollin, T. I., O'Connell, J. R., Sorkin, J. D., Agarwala, R., Schäffer, A. A., Streeten, E. A., King, T. M., Shuldiner, A. R., & Mitchell, B. D. (2006). Does having children extend life span? A genealogical study of parity and longevity in the Amish. *The Journals of Gerontology: Series A*, *61*(2), Article 2. https://doi.org/10.1093/gerona/61.2.190
Merkling, T., Blanchard, P., Chastel, O., Glauser, G., Vallat-Michel, A., Hatch, S. A., Danchin, E., & Helfenstein, F. (2017). Reproductive effort and oxidative stress: Effects of offspring sex and number on the physiological state of a long-lived bird. *Functional Ecology*, *31*(6), 1201–1209. https://doi.org/10.1111/1365-2435.12829
Monard, A.-M., Duncan, P., Fritz, H., & Feh, C. (1997). Variations in the birth sex ratio and neonatal mortality in a natural herd of horses. *Behavioral Ecology and Sociobiology*, *41*(4), 243–249. https://doi.org/10.1007/s002650050385
Myers, J. H. (1978). Sex ratio adjustment under food stress: Maximization of quality or numbers of offspring? *The American Naturalist*, *112*(984), 381–388. https://doi.org/10.1086/283280
Næss, Ø., Mortensen, L. H., Vikanes, Å., & Smith, G. D. (2017). Offspring sex and parental health and mortality. *Scientific Reports*, *7*(1), 5285. https://doi.org/10.1038/s41598-017-05161-y
Nath, D. C., & Land, K. C. (1994). Sex preference and third birth intervals in a traditional Indian society. *Journal of Biosocial Science*, *26*(3), 377–388. https://doi.org/10.1017/S0021932000021453
Nenko, I., Hayward, A. D., & Lummaa, V. (2014). The effect of socio-economic status and food availability on first birth interval in a pre-industrial human population. *Proceedings of the Royal Society B: Biological Sciences*, *281*(1775), Article 1775. https://doi.org/10.1098/rspb.2013.2319
Nichols, H. J., Fullard, K., & Amos, W. (2014). Costly sons do not lead to adaptive sex ratio adjustment in pilot whales, Globicephala melas. *Animal Behaviour*, *88*, 203–209. https://doi.org/10.1016/j.anbehav.2013.12.015
Nielsen, H. S., Mortensen, L., Nygaard, U., Schnor, O., Christiansen, O. B., & Andersen, A.-M. N. (2008). Brothers and reduction of the birth weight of later-born siblings. *American Journal of Epidemiology*, *167*(4), 480–484. https://doi.org/10.1093/aje/kwm330
Nishida, T. (1990). Demography and reproductive profiles. *University of Tokyo Press*, 63–98.
Noren, D. P. (2011). Estimated field metabolic rates and prey requirements of resident killer whales. *Marine Mammal Science*, *27*(1), 60–77. https://doi.org/10.1111/j.1748-7692.2010.00386.x

Ostner, J., Borries, C., Schülke, O., & Koenig, A. (2005). Sex allocation in a colobine monkey. *Ethology*, *111*(10), 924–939. https://doi.org/10.1111/j.1439-0310.2005.01102.x

Paul, A., & Kuester, J. (1990). Adaptive significance of sex ratio adjustment in semifree-ranging Barbary macaques (Macaca sylvanus) at Salem. *Behavioral Ecology and Sociobiology*, *27*(4). https://doi.org/10.1007/BF00164902

Paul, A., & Thommen, D. (1984). Timing of birth, female reproductive success and infant sex ratio in semifree-ranging barbary macaques (Macaca sylvanus). *Folia Primatologica*, *42*(1), 2–16. https://doi.org/10.1159/000156140

Pham-Kanter, G., & Goldman, N. (2012). Do sons reduce parental mortality? *J Epidemiol Community Health*, *66*(8), 710–715. https://doi.org/10.1136/jech.2010.123323

Plavcan, J. M. (2001). Sexual dimorphism in primate evolution. *American Journal of Physical Anthropology*, *116*(S33), 25–53. https://doi.org/10.1002/ajpa.10011

Pluháček, J., Bartoš, L., Doležalová, M., & Bartošová-Víchová, J. (2007). Sex of the foetus determines the time of weaning of the previous offspring of captive plains zebra (Equus burchelli). *Applied Animal Behaviour Science*, *105*(1–3), 192–204. https://doi.org/10.1016/j.applanim.2006.05.019

Pluháček, J., Bartošová, J., & Bartoš, L. (2011). Further evidence for sex differences in suckling behaviour of captive plains zebra foals. *Acta Ethologica*, *14*(2), 91–95. https://doi.org/10.1007/s10211-011-0091-z

Ralls, K. (1976). Mammals in which females are larger than males. *The Quarterly Review of Biology*, *51*(2), 245–276. https://doi.org/10.1086/409310

Rao, A. J., Ramesh, V., Ramachandra, S. G., Krishnamurthy, H. N., Ravindranath, N., & Moudgal, N. R. (1998). Growth and reproductive parameters of bonnet monkey (Macaca radiata). *Primates*, *39*(1), 97–107. https://doi.org/10.1007/BF02557748

Rickard, I. J. (2008). Offspring are lighter at birth and smaller in adulthood when born after a brother versus a sister in humans. *Evolution and Human Behavior*, *29*(3), 196–200. https://doi.org/10.1016/j.evolhumbehav.2008.01.006

Rivalan, P., Prévot-Julliard, A.-C., Choquet, R., Pradel, R., Jacquemin, B., & Girondot, M. (2005). Trade-off between current reproductive effort and delay to next reproduction in the leatherback sea turtle. *Oecologia*, *145*(4), 564–574. https://doi.org/10.1007/s00442-005-0159-4

Robbins, A. M., Robbins, M. M., & Fawcett, K. (2007). Maternal investment of the Virunga mountain gorillas. *Ethology*, *113*(3), 235–245. https://doi.org/10.1111/j.1439-0310.2006.01319.x

Robert, K. A., Schwanz, L. E., & Mills, H. R. (2010). Offspring sex varies with maternal investment ability: Empirical demonstration based on cross-fostering. *Biology Letters*, *6*(2), 242–245. https://doi.org/10.1098/rsbl.2009.0774

Rutkowska, J., Koskela, E., Mappes, T., & Speakman, J. R. (2011). A trade-off between current and future sex allocation revealed by maternal energy budget in a small mammal. *Proceedings of the Royal Society B: Biological Sciences*, *278*(1720), 2962–2969. https://doi.org/10.1098/rspb.2010.2654

Salogni, E., Galimberti, F., Sanvito, S., & Miller, E. H. (2019). Male and female pups of the highly sexually dimorphic northern elephant seal ( *Mirounga angustirostris* ) differ slightly in body size. *Canadian Journal of Zoology*, *97*(3), 241–250. https://doi.org/10.1139/cjz-2018-0220

Schino, G., Cozzolino, R., & Troisi, A. (1999). Social Rank and Sex-Biased Maternal Investment in Captive Japanese Macaques: Behavioural and Reproductive Data. *Folia Primatologica*, *70*(5), 254–263. https://doi.org/10.1159/000021704

Schwanz, L. E., & Robert, K. A. (2016). Costs of rearing the wrong sex: Cross-fostering to manipulate offspring sex in tammar wallabies. *PLOS ONE*, *11*(2), e0146011. https://doi.org/10.1371/journal.pone.0146011

Setchell, J. M., Lee, P. C., Wickings, E. J., & Dixson, A. F. (2002). Reproductive parameters and maternal investment in mandrills (Mandrillus sphinx). *International Journal of Primatology*, *23*(1), 51–68. https://doi.org/10.1023/A:1013245707228

Sievert, J., Karesh, W. B., & Sunde, V. (1991). Reproductive intervals in captive female western lowland gorillas with a comparison to wild mountain gorillas. *American Journal of Primatology*, *24*(3–4), 227–234. https://doi.org/10.1002/ajp.1350240308

Silk, J. B. (1983). Local resource competition and facultative adjustment of sex ratios in relation to competitive abilities. *The American Naturalist*, *121*(1), 56–66. https://doi.org/10.1086/284039

Silk, J. B. (1988). Maternal investment in captive bonnet macaques (Macaca radiata). *The American Naturalist*, *132*(1), 1–19. https://doi.org/10.1086/284834

Silk, J. B. (1990). Sources of variation in interbirth intervals among captive bonnet macaques ( *Macaca radiata* ). *American Journal of Physical Anthropology*, *82*(2), 213–230. https://doi.org/10.1002/ajpa.1330820210

Silk, J. B., Clark-Wheatley, C. B., Rodman, P. S., & Samuels, A. (1981). Differential reproductive success and facultative adjustment of sex ratios among captive female bonnet macaques (Macaca radiata). *Animal Behaviour*, *29*(4), 1106–1120. https://doi.org/10.1016/S0003-3472(81)80063-2

Simpson, A. E., & Simpson, M. J. A. (1985). Short-term consequences of different breeding histories for captive rhesus macaque mothers and young. *Behavioral Ecology and Sociobiology*, *18*(2), 83–89. https://doi.org/10.1007/BF00299036

Simpson, M. J. A., Simpson, A. E., Hooley, J., & Zunz, M. (1981). Infant-related influences on birth intervals in rhesus monkeys. *Nature*, *290*(5801), 49–51. https://doi.org/10.1038/290049a0

Small, M. F., & Smith, D. G. (1984). Sex differences in maternal investment by Macaca mulatta. *Behavioral Ecology and Sociobiology*, *14*(4), 313–314. https://doi.org/10.1007/BF00299503

Smuts, B., & Nicolson, N. (1989). Reproduction in wild female olive baboons. *American Journal of Primatology*, *19*(4), 229–246. https://doi.org/10.1002/ajp.1350190405

Stearns, S. C. (1992). The evolution of life histories. *Oxford University Press, 249*.

Strier, K. B. (1999). Predicting primate responses to “Stochastic” demographic events. *Primates*, *40*(1), 131–142. https://doi.org/10.1007/BF02557706

Strier, K. B., Mendes, S. L., & Santos, R. R. (2001). Timing of births in sympatric brown howler monkeys ( *Alouatta fusca clamitans* ) and northern muriquis ( *Brachyteles arachnoides hypoxanthus* ). *American Journal of Primatology*, *55*(2), 87–100. https://doi.org/10.1002/ajp.1042

Swenson, I., & Thang, N. M. (1993). Determinants of birth intervals in Vietnam: A hazard model analysis. *Journal of Tropical Pediatrics*, *39*(3), 163–167. https://doi.org/10.1093/tropej/39.3.163

Symington, M. M. (1987). Sex ratio and maternal rank in wild spider monkeys: When daughters disperse. *Behavioral Ecology and Sociobiology*, *20*(6), 421–425. https://doi.org/10.1007/BF00302985

Takahata, Y., Koyama, N., Huffman, M. A., Norikoshi, K., & Suzuki, H. (1995). Are daughters more costly to produce for Japanese macaque mothers?: Sex of the offspring and subsequent interbirth intervals. *Primates*, *36*(4), 571–574. https://doi.org/10.1007/BF02382877

Tamimi, R. M. (2003). Average energy intake among pregnant women carrying a boy compared with a girl. *BMJ*, *326*(7401), Article 7401. https://doi.org/10.1136/bmj.326.7401.1245

Tesfaw, L. M., Workie, D. L., & Dessie, Z. G. (2022). Discrepancies of recurrent birth intervals using longitudinal data analysis in Ethiopia: Interim EDHS 2019. *BMJ Open*, *12*(11), e066739. https://doi.org/10.1136/bmjopen-2022-066739

Toni, P., Forsyth, D. M., & Festa-Bianchet, M. (2021). Determinants of offspring sex in kangaroos: A test of multiple hypotheses. *Behavioral Ecology*, *32*(2), 297–305. https://doi.org/10.1093/beheco/araa131

Trillmich, F. (1986). Maternal investment and sex-allocation in the Galapagos fur seal, Arctocephalus galapagoensis. *Behavioral Ecology and Sociobiology*, *19*(3), 157–164. https://doi.org/10.1007/BF00300855

Trivers, R. L. (1972). Parental investment and sexual selection. In *Sexual selection and the descent of man* (1st edition, p. 44).

Trivers, R. L. (1985). *Social evolution*. Benjamin/Cummings Publ.

Trivers, R. L., & Willard, D. E. (1973). Natural selection of parental ability to vary the sex ratio of offspring. *Science*, *179*(4068), 90–92. https://doi.org/10.1126/science.179.4068.90

Trussell, J., Martin, L. G., Feldman, R., Palmore, J. A., Concepcion, M., & Abu Bakar, D. N. L. Bt. D. (1985). Determinants of birth-interval length in the Philippines, Malaysia, and Indonesia: A hazard-model analysis. *Demography*, *22*(2), 145–168. https://doi.org/10.2307/2061175

Tu, P. (1991). Birth spacing patterns and correlates in Shaanxi, China. *Studies in Family Planning*, *22*(4), 255. https://doi.org/10.2307/1966481

Van De Putte, B., Matthijs, K., & Vlietinck, R. (2004). A social component in the negative effect of sons on maternal longevity in pre-industrial humans. *Journal of Biosocial Science*, *36*(3), Article 3. https://doi.org/10.1017/S0021932003006266

Van Noordwijk, A. J., & de Jong, G. (1986). Acquisition and allocation of resources: Their influence on variation in life history tactics. *The American Naturalist*, *128*(1), Article 1. https://doi.org/10.1086/284547

Van Schaik, C. P., Netto, W. J., Van Amerongen, A. J. J., & Westland, H. (1989). Social rank and sex ratio of captive long-tailed macaque females ( *Macaca fascicularis* ). *American Journal of Primatology*, *19*(3), 147–161. https://doi.org/10.1002/ajp.1350190303

Verme, L. J. (1989). Maternal investment in white-tailed deer. *Journal of Mammalogy*, *70*(2), 438–442. https://doi.org/10.2307/1381538

Weimerskirch, H., Barbraud, C., & Lys, P. (2000). Sex differences in parental investment and chick growth in wandering albatrosses: Fitness consequences. *Ecology*, *81*(2), 309–318. https://doi.org/10.1890/0012-9658(2000)081[0309:SDIPIA]2.0.CO;2

Weiss, M. N., Ellis, S., Franks, D. W., Nielsen, M. L. K., Cant, M. A., Johnstone, R. A., Ellifrit, D. K., Balcomb, K. C., & Croft, D. P. (2023). Costly lifetime maternal investment in killer whales. *Current Biology*, *33*(4), 744-748.e3. https://doi.org/10.1016/j.cub.2022.12.057

White, A. M., Swaisgood, R. R., & Czekala, N. (2007). Differential Investment in Sons and Daughters: Do White Rhinoceros Mothers Favor Sons? *Journal of Mammalogy*, *88*(3), 632–638. https://doi.org/10.1644/06-MAMM-A-180R2.1

Wilkinson, I. S., & Van Aarde, R. J. (2001). Investment in sons and daughters by elephant seals, Mirounga leonina, at Marion Island. *Marine Mammal Science*, *17*(4), 873–887. https://doi.org/10.1111/j.1748-7692.2001.tb01303.x

Wolff, J. O. (1988). Maternal investment and sex ratio adjustment in American bison calves. *Behavioral Ecology and Sociobiology*, *23*(2), 127–133. https://doi.org/10.1007/BF00299896

Wright, B. M., Stredulinsky, E. H., Ellis, G. M., & Ford, J. K. B. (2016). Kin-directed food sharing promotes lifetime natal philopatry of both sexes in a population of fish-eating killer whales, Orcinus orca. *Animal Behaviour*, *115*, 81–95. https://doi.org/10.1016/j.anbehav.2016.02.025

Yi, B., Wang, S., Sun, T., Liu, R., Lawes, M. J., Yang, L., Liu, X., Li, Y., Huang, C., Zhou, Q., & Fan, P. (2023). Maternal parity influences the birth sex ratio and birth interval of captive Francois' langur (Trachypithecus francoisi). *Behavioral Ecology and Sociobiology*, *77*(12), 140. https://doi.org/10.1007/s00265-023-03408-5

Zhao, Q., Jin, T., Wang, D., Qin, D., Yin, L., Ran, W., & Pan, W. (2009). Lack of sex-biased maternal investment in spite of a skewed birth sex ratio in white-headed langurs ( *Trachypithecus leucocephalus* ). *Ethology*, *115*(3), 280–286. https://doi.org/10.1111/j.1439-0310.2008.01609.x

Ziomkiewicz, A., Sancilio, A., Galbarczyk, A., Klimek, M., Jasienska, G., & Bribiescas, R. G. (2016). Evidence for the Cost of Reproduction in Humans: High Lifetime Reproductive Effort Is Associated with Greater Oxidative Stress in Post-Menopausal Women. *PLOS ONE*, *11*(1), Article 1. https://doi.org/10.1371/journal.pone.0145753

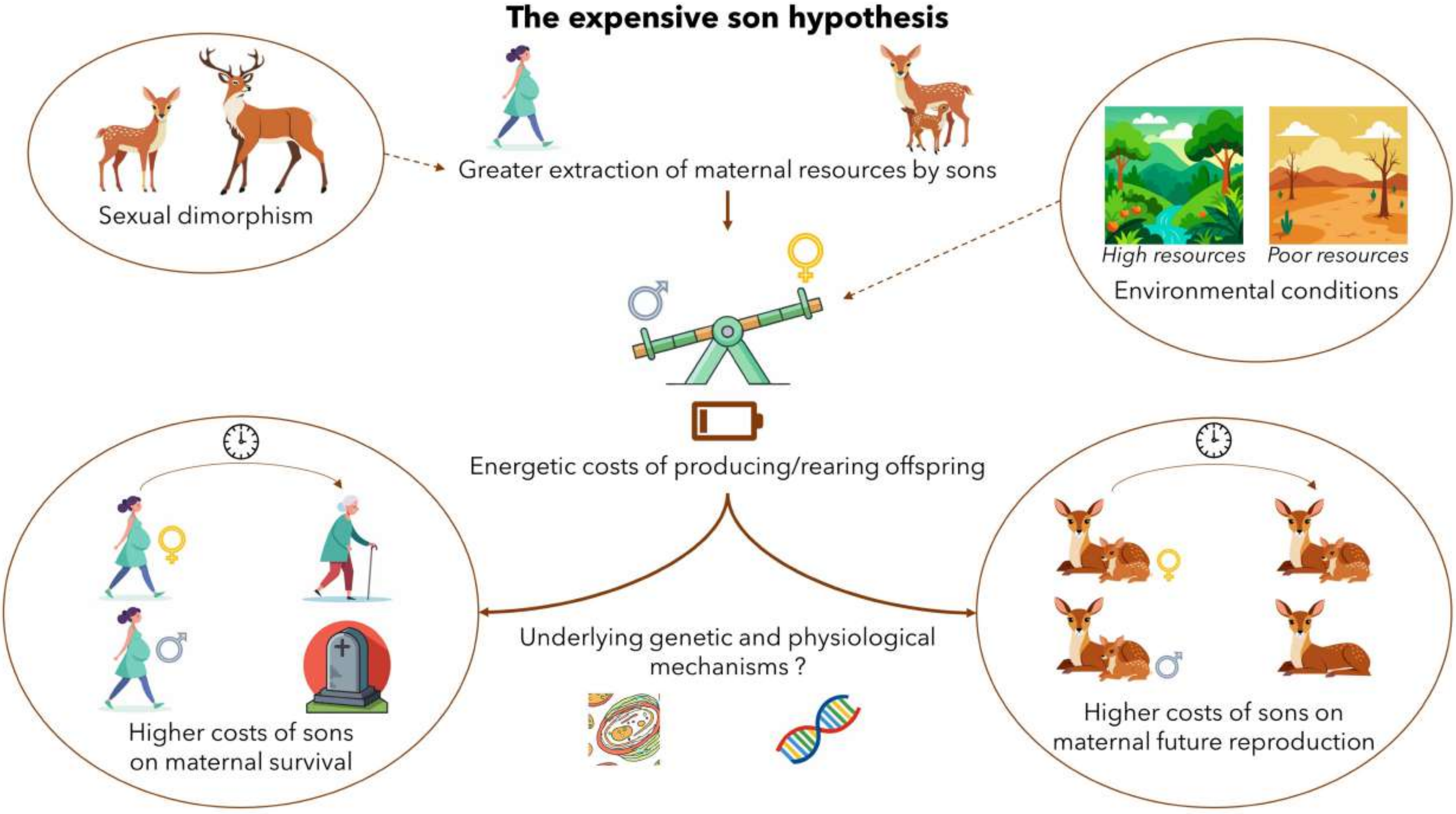
The expensive son hypothesis
Sexual dimorphism
Greater extraction of maternal resources by sons
High resources
Poor resources
Environmental conditions
Energetic costs of producing/rearing offspring
Underlying genetic and physiological mechanisms ?
Higher costs of sons on maternal survival
Higher costs of sons on maternal future reproduction

**Table 1**. A review of studies testing the differential costs of sons over daughters on maternal future reproduction. The literature search protocol used is available in electronic supplementary material. A support for the higher costs of sons was considered when the p-value was below the threshold of 5%, none of studies have used Bayesian methods. The SSD index was calculated by dividing the sum of the mass of males and females by the mass of males. When the mass was not available we used the body size. A positive value means that the SSD is biased towards males while a negative means the SSD is biased towards females. Regarding the population type, “semi-captive” refers to lightly managed populations (supplemental feeding).

| **Species** | **Metrics** | **Higher costs of sons over daughters** | **Factors modulating differential costs of offspring sex** | **Population** | **Population type** | **Weaning SSD** | **Adult SSD** | **References** |
|---|---|---|---|---|---|---|---|---|
| Orca (*Orcinus orca*) | Probability to produce a viable calf for mothers having surviving sons | Yes | Not tested | Southern resident (Canada & USA) | Wild | NA* | 0.247 | (Weiss et al., 2023) |
| Southern elephant seal (*Mirounga leonina*) | Probability of giving birth according to the sex of offspring produced the previous season | No | Size of females tested but no effect | Marion Island (South Africa) | Wild | 0.139 | 0.800 | (Wilkinson & Van Aarde, 2001) |
| Northern elephant seal (*Mirounga angustirostris*) | Probability of giving birth according to the sex of offspring | No | Not tested | California (USA) | Wild | 0.076 | 0.751 | (Le Bœuf et al., 1989) |
| | weaned the previous season | | | | | | | |
| Long-finned pilot whale (*Globicephala melas*) | Probability of giving birth according to the sex of offspring from the last reproduction being raised | Yes | Not tested | Faroe Islands | Wild | NA | NA | (Nichols et al., 2014) |
| Antarctic fur seal (*Arctocephalus gazella*) | Probability of giving birth according to the sex of offspring produced the previous season | No | Not tested | Bird Island (South Georgia) | Wild | 0.092 | 0.750 | (Lunn & Arnould, 1997) |
| Galapagos fur seal (*Arctocephalus galapagoensis*) | Probability of giving birth according to the sex of offspring weaned the previous season | No | Not tested | Fernandina Island (Ecuador) | Wild | NA | 0.555 | (Trillmich, 1986) |
| Red deer (*Cervus elaphus*) | Probability of giving birth according to the sex of offspring produced the previous season | Yes | Survival of the offspring tested but no effect | Isle of Rum (Scotland) | Wild | NA | NA | (Froy et al., 2016) |
| | Probability of giving birth according to the sex of offspring | No | Costs of rearing sons occur only in low-ranking females | | | | | (Gomendio et al., 1990) |
| | weaned the previous season | | | | | | | |
| | Birth interval* | Yes | Not tested | | | | | (Clutton-Brock et al., 1981) |
| | Probability of giving birth according to the sex of offspring weaned the previous season | Yes | | | | | | |
| Red deer (*Cervus elaphus*) | Probability of giving birth according to the sex of offspring weaned the previous season | No | Not tested | The Petite Pierre National Reserve (France) | Wild | 0.087 | NA | (Bonenfant et al., 2003) |
| Bighorn sheep (*Ovis canadensis*) | Age-specific decline in fecundity with age according to the sex ratio of offspring weaned in early life | Yes | Not tested | Ram Mountain (Canada) | Wild | NA | 0.170 | (Douhard et al., 2020) |
| | Survival of the lamb according to the sex of the offspring weaned the previous season | Yes | Costs of rearing sons are greater at high population density | | | | | (Berube et al., 1996) |
| | Lamb summer mass gain | Yes | Not tested | | | | | (Martin & Festa-Bianchet, 2010) |
| Bighorn sheep (*Ovis canadensis*) | Duration to return to oestrus according to the sex of the offspring produced the previous season | Yes | Age of ewe tested but no effect | Montana (USA) | Semi-captive | NA | NA | (Hogg et al., 1992) |
| Bighorn sheep (*Ovis canadensis*) | Probability of giving birth according to the sex of offspring weaned the previous season | No | Not tested | Sheep River (Canada) | Wild | 0.098 | 0.335 | (Festa-Bianchet, 1989) |
| Soay sheep (*Ovis aries*) | Birth weight of lamb according to the sex of twins previously birthed | Yes | Not tested | St Kilda archipelago (Scotland) | Wild | 0.121 | 0.198 | (Korsten et al., 2009) |
| Fallow deer (*Dama dama*) | Birth interval* | Not tested | Costs of rearing sons occur in mothers aged >3 years only | Tovetorp (Sweden) | Captive | -0.019 | NA | (Birgersson, 1998) |
| White-tailed deer (*Odocoileus virginianus*) | Birth interval* | No | Age of mother and litter size (twins or singletons) tested but no effect | Shingleton (USA) | Captive | 0.096 | NA | (Verme, 1989) |
| Mountain goats (*Oreamnos americanus*) | Probability of giving birth according to the sex of offspring | No | Mass of the offspring tested but no effect | Caw Ridge (Canada) | Wild | 0.096 | 0.310 | (Hamel et al., 2011) |
| | produced the previous season | | | | | | | |
| | Probability of giving birth according to the sex of yearling observed the previous season | Yes | Not tested | | | | | (Charest Castro et al., 2018) |
| American bison (*Bison bison*) | Probability of giving birth according to the sex of offspring produced the previous season | Yes | Not tested | Montana (USA) | Semi-captive | NA | NA | (Wolff, 1988) |
| American bison (*Bison bison*) | Birth interval* | No | Not tested | South Dakota (USA) | Semi-captive | NA | NA | (Green & Rothstein, 1991) |
| | Probability of giving birth according to the sex of offspring produced the previous season | No | | | | | | |
| Giraffes (*Giraffa camelopardalis*) | Birth interval* | No | Identity of the mother tested but no effect | San Diego zoo (USA) | Captive | NA | 0.349 | (Bercovitch et al., 2004) |
| Hariana cattle (*Bos taurus*) | Birth interval* | No | Not tested | Unspecified | NA | NA | NA | (Dhillon et al., 1970) |
| Friesian cattle (*Bos Taurus*) | Birth interval* | No | Not tested | Cairo (Egypt) | Captive | NA | NA | (Dessouky & Rakha, 1961) |
| Cuvier's gazelle (*Gazella cuvieri*) | Probability of giving birth according to the sex ratio of offspring produced the previous seasons | No | Not tested | Almeria (Spain) | Captive | NA | NA | (Alados & Escós, 1994) |
| | Birth interval* | No | | | | | | |
| Humans | Birth weight of child according to the sex of the previous child | Yes | Birth order and birth interval tested but no effect | Glasgow (Scotland) | NA | NA | NA | (Rickard, 2008) |
| Humans | Birth interval* | No; higher costs of daughters | Not tested | Hutterite sect | NA | NA | NA | (Margulis et al., 1993) |
| Humans | Birth interval* | No | Birth interval longer after the birth of male co-twins than female co-twins | Finland | NA | NA | NA | (Lummaa, 2001) |
| | Probability of giving birth according to the sex of the last offspring produced | No | Mothers are more like to cease reproducing after the birth of male co-twins than female co-twins | | | | | |
| Humans | Birth interval* | Yes | Not tested | Kenya | NA | NA | NA | (Mace & Sear, 1997) |
| Humans | Birth interval* | Yes | Not tested | Ethiopia | NA | NA | NA | (Tesfaw et al., 2022) |
| Humans | Birth weight of boys according to the sex of previous children | Yes | Not tested | Colleges and universities in USA and Canada | NA | NA | NA | (Blanchard & Ellis, 2001) |
| | Birth weight of daughters according to the sex of previous children | No | | | | | | |
| Humans | Birth weight of boys according to the sex of previous children | Yes | Not tested | Quebec (Canada) | NA | NA | NA | (Côté et al., 2003) |
| | Birth weight of daughters according to the sex of previous children | No | | | | | | |
| Humans | Birth weight of children according to the sex of the first child | Yes | Not tested | Norway | NA | NA | 0.131 | (Magnus et al., 1985) |
| Humans | Birth weight of child according to the sex of the previous child | Yes | Not tested | Denmark | NA | NA | NA | (Nielsen et al., 2008) |
| Humans | Birth interval* | No | Not tested | Vietnam | NA | NA | NA | (Swenson & Thang, 1993) |
| Humans | Length of the third birth interval* as a | No | Not tested | Assam (India) | NA | NA | NA | (Nath & Land, 1994) |
| | function of the sex ratio of the two-preceding offspring | | | | | | | |
| Humans | Birth interval* | No | Not tested | Philippines | NA | NA | NA | (Trussell et al., 1985) |
| | Birth interval* | No | Not tested | Malaysia | NA | NA | NA | |
| | Birth interval* | No | Not tested | Indonesia | NA | NA | NA | |
| Humans | Birth interval* as a function of the sex ratio of preceding offspring | Yes | Not tested | Bangladesh | NA | NA | NA | (Bairagi & Langsten, 1986) |
| Humans | Probability of giving birth according to the sex ratio of previous offspring produced | Yes | Not tested | Shaanxi (China) | NA | NA | NA | (Tu, 1991) |
| | Birth interval* | Yes | | | | | | |
| Humans | Birth weight of child according to the sex of the previous child | Yes | Not tested | Sweden | NA | NA | NA | (Jelenkovic et al., 2014) |
| | Birth interval* | No | | | | | | |
| Rhesus macaques (*Macaca mulatta*) | Probability of giving birth according to the sex of offspring | No | Costs of rearing sons occur in dominant mothers, whereas costs of rearing daughters | Madingley (UK) | Captive | -0.065 | NA | (Gomendio et al., 1990) |
| | raised the previous season | | occur in subordinate mothers | | | | | |
| | Birth interval* | No | Birth interval longer after the birth of a daughter for low-ranking females | | | | | (Gomendio, 1990) |
| | Probability of giving birth according to the sex of offspring raised the previous season | No | A lower probability to reproduce after raising a daughter for low-ranking females | | | | | |
| | Birth interval* | No; higher costs of daughters | Not tested | | | | | (M. J. A. Simpson et al., 1981) |
| | Birth interval* | No | Birth interval longer after raising a daughter for high-ranking females | | | | | (A. E. Simpson & Simpson, 1985) |
| Rhesus macaques (*Macaca mulatta*) | Birth interval* | No | Not tested | California (USA) | Captive | NA | 0.360 | (Small & Smith, 1984) |
| | Probability of produce a viable offspring according to the sex of offspring produced the previous season | Not tested | Maternal status (primiparous or multiparous) tested but no effect | | | | | (Hinde, 2009) |
| | Birth interval* | No | Maternal status (primiparous or multiparous) tested but no effect | | | | | |
| | Probability of conceive according to the sex of offspring produced the previous season | Not tested | Primiparous mothers of sons more likely to conceive during the next breeding season | | | | | |
| Rhesus macaques (Macaca mulatta) | Birth interval* | Yes | Not tested | Cayo Santiago (Puerto Rico) | Semi-captive | 0.095 | 0.193 | (Berman, 1988) |
| Rhesus macaques (*Macaca mulatta*) | Birth interval* | No; higher costs of daughters | Maternal rank tested but no effect | Lawrenceville (USA) | Captive | NA | NA | (Maestripieri, 2001) |
| Japanese macaques (*Macaca fuscata*) | Probability of giving birth according to the sex of offspring produced the previous season | Not tested | A lower probability to reproduce after raising a daughter for low-ranking females | Takasakiyama Natural zoo (Japan) | Captive | -0.016 | 0.332 | (Kurita et al., 2012) |
| Japanese macaques (*Macaca fuscata*) | Birth interval* | No | Maternal rank tested but no effect | Rome zoo (Italia) | Captive | NA | NA | (Schino et al., 1999) |
| Japanese macaques (*Macaca fuscata*) | Birth interval* | No; higher costs of daughters | Birth interval longer after the birth of a daughter | Kyoto (Japan) | Captive | 0.098 | 0.268 | (Takahata et al., 1995) |
| | | | in low-ranking females only | | | | | |
| Bonnet macaques (*Macaca radiata*) | Birth interval* | No | Survival of the offspring opening the interval tested but no effect | California (USA) | Captive | NA | NA | (Silk, 1988) |
| | Birth interval* | No | Maternal rank tested but no effect | | | | | (Silk et al., 1981) |
| | Birth interval* | No | Survival of the offspring opening the interval tested but no effect | | | | | (Silk, 1990) |
| Barbary macaques (*Macaca sylvanus*) | Birth interval* | No | Maternal rank tested but no effect | Salem (USA) | Semi-captive | NA | NA | (Paul & Kuester, 1990) |
| | Birth interval* | Not tested | Parity tested but not effect | | | | | (Paul & Thommen, 1984) |
| Pigtail macaques (*Macaca nemestrina leonine*) | Birth interval* | No; higher costs of daughters | Maternal rank and age tested but no effect | Lawrenceville (USA) | Captive | NA | NA | (Maestripieri, 2002) |
| Long-tailed macaques (*Macaca fascicularis*) | Birth interval* | No; higher costs of daughters | Birth interval longer after the birth of a daughter in high-ranking females | Arnhem (Netherlands) | Captive | NA | NA | (Van Schaik et al., 1989) |
| Mandrill (*Mandrillus sphinx*) | Birth interval* | No | Parity and social rank tested but no effect | Franceville (Gabon) | Semi-captive | 0.021 | 0.708 | (Setchell et al., 2002) |
| Chimpanzee (*Pan troglodytes*) | Birth interval* | Yes | Not tested | Kibale National Park (Uganda) | Wild | NA | NA | (Bădescu et al., 2022) |
| Chimpanzee (*Pan troglodytes*) | Birth interval* | No | Survival of the offspring opening the birth interval tested but no effect | Mahale Mountains (Tanzania) | Wild | NA | 0.162 | (Nishida, 1990) |
| Chimpanzee (*Pan troglodytes*) | Birth interval* | No | Birth interval longer after the birth of a son in high-ranking females; birth interval longer after the birth of a daughter in low-ranking females | Taï tropical rainforest (Côte d'Ivoire) | Wild | NA | NA | (Boesch, 1997) |
| Mountain gorillas (*Gorilla beringei beringei*) | Birth interval* | No | Birth interval longer after the birth of a son in high ranking females; Birth interval longer after the birth of a daughter in low ranking females | Karisoke Research Center (Rwanda) | Semi-captive | NA | NA | (Robbins et al., 2007) |
| Western lowland gorillas (*Gorilla gorilla gorilla*) | Birth interval* | No | Not tested | Various institutions in North America and Europe | Captive | NA | NA | (Sievert et al., 1991) |
| Red howler monkey | Birth interval* | No | Habitat and maternal age tested but no effect | Hato Masaguaral (Venezuela) | Wild | 0 | 0.279 | (Crockett & Rudran, 1987) |
| (*Alouatta seniculus*) | | | | | | | | |
| Howler monkeys (*Alouatta fusca clamitans*) | Birth interval* | No | Not tested | Estação Biológica de Caratinga (Brazil) | Wild | NA | NA | (Strier et al., 2001) |
| Spider monkey (*Ateles paniscus*) | Birth interval* | Not tested | Birth interval longer after the birth of a son in high ranking females | Manu National Park (Peru) | Wild | NA | NA | (Symington, 1987) |
| Olive baboons (*Papio cynocephalus anubis*) | Birth interval* | No | Not tested | Gilgil (Kenya) | Wild | 0 | 0.480 | (Smuts & Nicolson, 1989) |
| Francois' langur (*Trachypithecus francoisi*) | Birth interval* | No; higher costs of daughters | Maternal age and parity tested but no effect | Various zoos in China | Captive | NA | NA | (Yi et al., 2023) |
| Hanuman langurs (*Semnopithecus entellus*) | Birth interval* | Yes | Not tested | Ramnagar (Nepal) | Wild | NA | NA | (Ostner et al., 2005) |
| White-headed langurs (*Trachypithecus leucocephalus*) | Birth interval* | No | Not tested | Nongguan Natural Reserve (China) | Wild | NA | NA | (Zhao et al., 2009) |
| Muriquis (*Brachyteles arachnoides*) | Birth interval* | No | Not tested | Estação Biológica de Caratinga (Brazil) | Wild | 0 | 0 | (Strier, 1999) |
| Plains zebra (*Equus quagga*) | Birth interval* | Yes | Not tested | Hwange National Park (Zimbabwe) | Wild | 0 | 0 | (Barnier et al., 2012) |
| Plains zebra (*Equus burchelli*) | Weaning age according to the sex of the foetus | Yes | Not tested | Dvu˚r Kra´love´ Zoo (Czech Republic) | Captive | 0 | 0 | (Pluháček et al., 2007) |
| White rhinoceros (*Ceratotherium simum*) | Birth interval* | No | Not tested | Hluhluwe- iMfolozi Game Park (South Africa) | Wild | NA | NA | (White et al., 2007) |
| Horse (*Equus caballus*) | Birth interval* | Not tested | Birth interval longer after the birth of a son in poor years | Camargue (France) | Wild | NA | NA | (Monard et al., 1997) |
| | Probability of giving birth according to the sex of the offspring weaned during the year | No | Not tested | | | | | |
| Horse (*Equus caballus*) | Probability of giving birth according to the sex of the offspring produced the previous season | No | Not tested | Central North Island (New Zealand) | Wild | NA | NA | (Cameron & Linklater, 2000) |
| | Birth interval* | No | Birth interval longer after the birth of a daughter in poor conditions females | | | | | |
| Tammar wallabies (*Macropus eugenii*) | Probability of giving birth according to the sex of the offspring reared in a case of cross-fostering | No | Not tested | Garden Island (Australia) | Semi-captive | NA | NA | (Schwanz & Robert, 2016) |
| Tammar wallabies (*Macropus eugenii*) | Probability to wean a cross-fostered offspring according to the sex of previous offspring birthed | No | Not tested | Tutanning Nature Reserve (Australia) | Wild | NA | NA | (Robert et al., 2010) |
| Eastern gray kangaroos (*Macropus giganteus*) | Probability of weaning an offspring as a function of the sex of the previous offspring produced | Yes | Maternal mass tested but no effect | Wilsons Promontory National Park (Australia) | Wild | NA | NA | (Toni et al., 2021) |
| African elephant (*Loxodonta africana*) | Birth interval* | Yes | Not tested | Amboseli National Park (Kenya) | Wild | NA | 0.267 (using height) | (Lee & Moss, 1986) |
| Bank vole (*Myodes glareolus*) | Probability of giving birth according to the sex ratio of the previous litter born | No | Previous litter size tested but no effect | Konnevesi (Finland) | Captive | 0.006 | 0.099 | (Koskela et al., 2009) |
| | Birth weight of males according to the sex ratio of the previous litter weaned | No | Not tested | | | | | (Rutkowska et al., 2011) |
| | Birth weight of females according to the sex ratio of the previous litter weaned | Yes | Not tested | | | | | |
| Columbian ground squirrel (*Urocitellus columbianus*) | Litter size weaned according to the litter sex ratio at birth | No | Not tested | Sheep River (Canada) | Wild | 0.036 | NA | (Barra et al., 2021) |
| | Litter size weaned according to the litter sex ratio at weaning | No | Not tested | | | | | |
| Mongolian gerbils (*Meriones unguiculatus*) | Birth interval* | Yes | Not tested | McMaster colony (Canada) | Captive | 0.008 | 0.166 | (Clark et al., 1990) |

*Birth interval refers here to the method used to test whether the sex of the offspring opening the interval influences the duration until the next birth. Although some studies did not take the survival of the child opening the interval into account, others only considered birth intervals in which offspring survived to a given age (ex: weaning, 1 year…). NA means the information is not available for the population.

**Table 2.** A review of studies testing the differential costs of sons and daughters on maternal survival. The literature search protocol is available in electronic supplementary material. A support for the higher costs of sons was considered when the p-value was below the threshold of 5%, none of studies have used Bayesian methods. The SSD index was calculated by dividing the sum of the mass of males and females by the mass of males. When the mass was not available we used the body size. A positive value means that the SSD is biased towards males while a negative means the SSD is biased towards females. Regarding the population type, "semi-captive" refers to lightly managed populations (supplemental feeding, reduced area, demography controlled with hunting). A human population is considered modern if the survey period is after 1760, else it is considered pre-industrial.

| **Species** | **Metrics** | **Higher costs of sons over daughters** | **Factors modulating differential costs of offspring sex** | **Population** | **Population type** | **Weaning SSD** | **Adult SSD** | **References** |
|---|---|---|---|---|---|---|---|---|
| Northern elephant seal (*Mirounga angustirostris*) | Survival probability of the mother from t to t+1 according to sex of the offspring produced at t | No | Not tested | California (USA) | Wild | 0.076 | 0.751 | (Le Bœuf et al., 1989) |
| Red deer (*Cervus elaphus*) | Survival probability of the mother from t to t+1 according to sex of the | Not tested | Yes, costs of sons occur only in low-ranking females | Isle of Rum (Scotland) | Wild | NA* | NA | (Gomendio et al., 1990) |
| | offspring weaned at t | | | | | | | |
| | Survival probability of the mother from t to t+1according to sex of the offspring produced at t | Yes | Survival of the offspring tested but no effect | | | | | (Froy et al., 2016) |
| | Lifespan of mothers surviving at least to 10 years of age as a function of the proportion of sons born and weaned between 3 and 9 years of age | No | Number of offspring born and weaned between 3 and 9 years of age but no effect | | | | | (Douhard et al., 2019) |
| Bighorn sheep (*Ovis canadensis*) | Survival probability of the mother from t to t+1 according to sex of the offspring weaned at t | No | Not tested | Ram Mountain (Canada) | Wild | NA | 0.170 | (Berube et al., 1996) |
| | Lifespan of mothers surviving at least to 8 years of age as a function of the proportion of sons born and weaned | No | Number of offspring born and weaned between 2 and 7 years of age but no effect | | | | | (Douhard et al., 2019) |
| | between 2 and 7 years of age | | | | | | | |
| Bighorn sheep (*Ovis canadensis*) | Survival probability of the mother from t to t+1 according to sex of the offspring weaned at t | No | Not tested | Sheep River (Canada) | Wild | 0.098 | 0.335 | (Festa-Bianchet, 1989) |
| Soay sheep (*Ovis aries*) | Lifespan of females surviving at least to 7 years of age as a function of the proportion of sons born and weaned between 1 and 6 years of age | No | Number of offspring born and weaned between 1 and 6 years of age but no effect | Hirta (Scotland) | Wild | 0.121 | 0.198 | (Douhard et al., 2019) |
| Mountain goats (*Oreamnos americanus*) | Survival probability of the mother from t to t+1 according to sex of the offspring produced at t | No | Offspring mass tested but no effect | Caw Ridge (Canada) | Wild | 0.096 | 0.310 | (Hamel et al., 2011) |
| | Lifespan of mothers surviving at least to 10 years of age as a function of the proportion of sons born and weaned | No | Number of offspring born and weaned between 3 and 9 years of age but no effect | | | | | (Douhard et al., 2019) |
| | between 3 and 9 years of age | | | | | | | |
| Columbian ground squirrel (*Urocitellus columbianus*) | Survival probability of the mother from t to t+1 according to sex of the offspring produced and weaned at t | No | No tested | Sheep River (Canada) | Wild | 0.036 | NA | (Barra et al., 2021) |
| Golden-mantled ground squirrel (*Callospermophilus lateralis*) | Survival probability of the mother from t to t+1 as a function of litter sex ratio produced at t | No | No tested | Colorado (USA) | Wild | NA | NA | (Kanaziz et al., 2022) |
| Bonnet macaques (*Macaca radiata*) | Maternal mortality just after birth according to the sex of the offspring produced at t | No | Not tested | California (USA) | Captive | NA | NA | (Silk, 1988) |
| Humans | Lifespan of women surviving at least to 50 years of age as a function of the proportion of sons born | No | Not tested | Quebec (Canada) | Pre-industrial | NA | NA | (Beise & Voland, 2002) |
| | | No | Not tested | Germany | Mainly modern | NA | NA | |
| Humans | Lifespan of women surviving at least to 50 years of age as a function of the | Yes | Not tested | Sami (Finland) | Mainly pre-industrial | NA | NA | (Helle et al., 2002) |
| | proportion of sons born | | | | | | | |
| Humans | Lifespan of women surviving at least to 50 years of age as a function of the proportion of sons born | No | Not tested | Amish (USA) | Modern | NA | NA | (McArdle et al., 2006) |
| Humans | Lifespan of women surviving at least to 50 years of age as a function of the proportion of sons born and surviving until 5 years | No | Yes, costs of sons occur only in the lowest social class | Moerzeke (Belgium) | Mainly modern | NA | NA | (Van De Putte et al., 2004) |
| Humans | Lifespan of women surviving at least to 45 years of age as a function of the proportion of sons born | No | Yes, costs of sons occur only when adjusting for the number of surviving sons. Socio-economic status tested but no effect. | Matlab (Bangladesh) | Modern | NA | NA | (Hurt et al., 2006) |
| Humans | Lifespan of women as well lifespan of those surviving at least to 50 years of age as a function of the proportion of sons born | No | Socio-economic status tested but no effect | Skelleftea (Sweden) | Pre-industrial | NA | NA | (Cesarini et al., 2007) |
| Humans | Lifespan of women surviving at least to 50 years of age as a function of the proportion of sons born and surviving to 15 years | Yes | Socio-economic status tested but no effect | Utah (USA) | Modern | NA | NA | (Harrell et al., 2008) |
| Humans | Lifespan of women surviving at least to 45 years of age as a function of the proportion of sons born | No | Not tested | Poland | Modern | NA | NA | (Jasienska et al., 2006) |
| Humans | Lifespan of women surviving at least to 50 years of age as a function of the proportion of sons born | No | Not tested | Sami (Sweden) | Mainly modern | NA | NA | (Cesarini et al., 2009) |
| Humans | Lifespan of women surviving at least to 50 years of age as a function of the proportion of sons born | Yes | Not tested | Sami (Finland and Sweden) | Mainly pre-industrial | NA | NA | (Helle et al., 2010) |
| Humans | Lifespan of women as a function of the proportion of sons born and surviving to adulthood | Yes, when sons were considered at birth | Socio-economic status tested but no effect | Finland | Mainly modern | NA | NA | (Helle & Lummaa, 2013) |
| Humans | Lifespan of women surviving at least to 65 years of age as a function of the proportion of surviving sons | No, higher costs of daughters | Not tested | Korea | Modern | NA | NA | (Kim, 2022) |
| Humans | Mortality risk of women surviving at least to 50 years of age as a function of the sex of the two first child | Yes | Not tested | Norway | Modern | NA | NA | (Næss et al., 2017) |
| Humans | Mortality risk of women aged less than 40 years of age as a function of the proportion of sons | No | Yes, costs of sons occur only when comparing mothers of four sons with those of four daughters | Norway | Modern | NA | NA | (Christiansen, 2014) |
| Humans | Mortality risk of women surviving at least to 60 years of age as a function of the proportion of sons born | No | Not tested | China | Modern | NA | NA | (Pham-Kanter & Goldman, 2012) |
| | Mortality risk of women surviving at least to 65 years of age as a | No | Not tested | Taiwan | Modern | NA | NA | |
| | function of the proportion of sons born | | | | | | | |
| Humans | Mortality risk of women as a function of the proportion of sons born | No | Not tested | USA | Modern | NA | NA | (Grandi et al., 2023) |
| Humans | Mortality risk of women surviving at least 45 years of age as a function of the proportion of sons born, of sons born, weaned and raised to adulthood | No | Yes, costs of sons occur in women having high infant mortality rates | Quebec (Canada) | Pre-industrial | NA | NA | (Invernizzi et al., 2024) |

*NA means the information is not available for the population

**Supplementary material 1.** Prisma diagram of the literature research protocol used for the Table 1

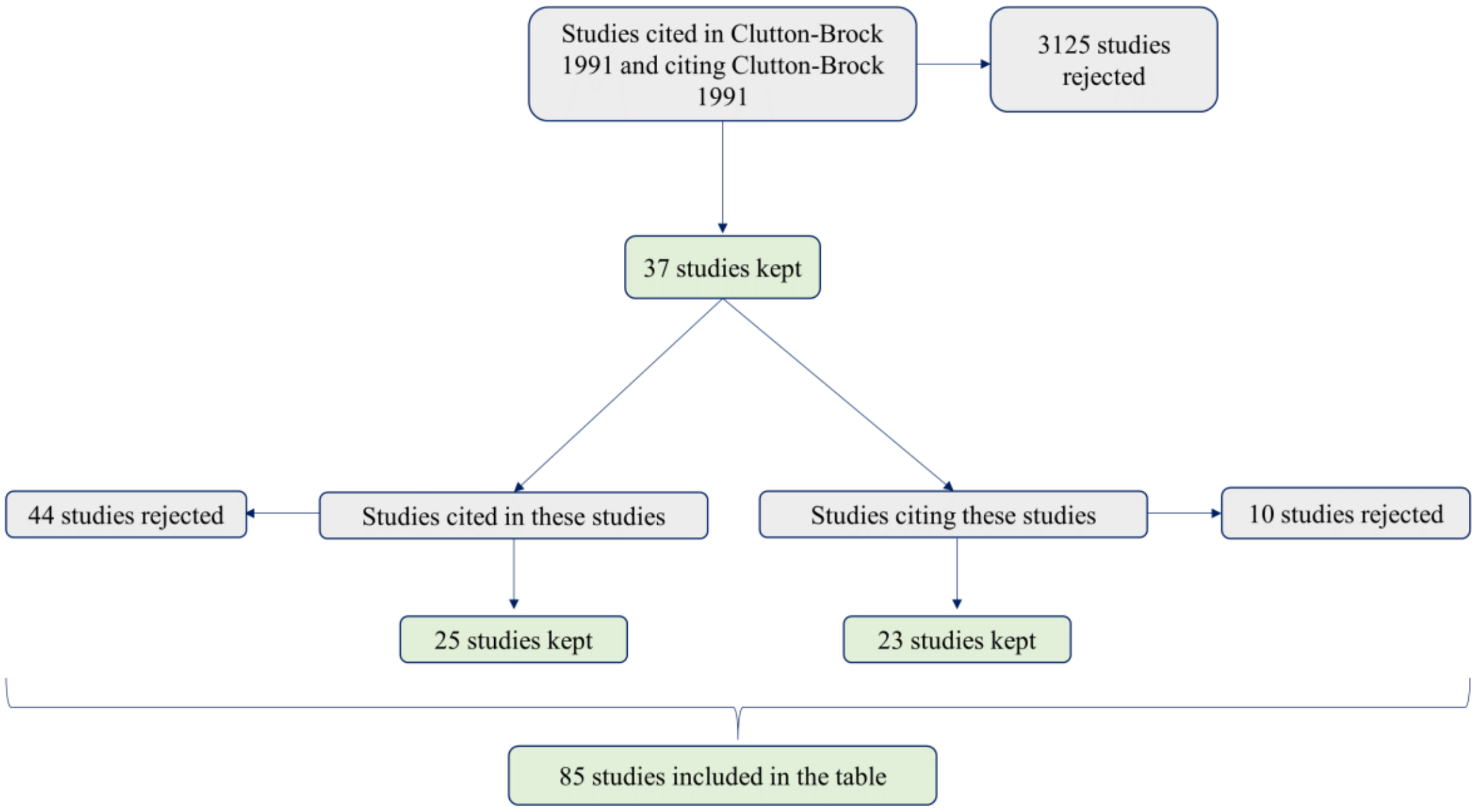

**Supplementary material 2.** Prisma diagram of the literature research protocol used for the Table 2

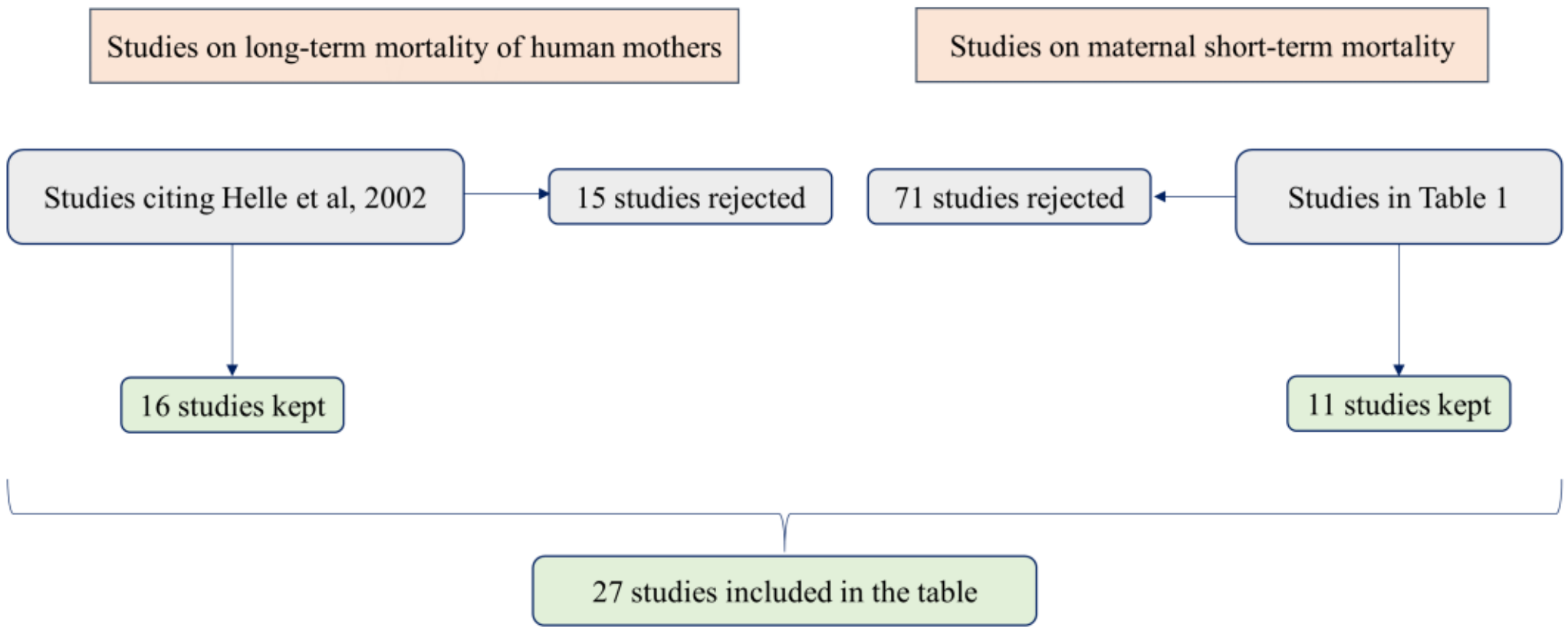

**Supplementary material 3: Research protocol for information about SSD in Table 1 & 2.**

Firstly, we looked to see if information on weight or height at adulthood and at weaning was available in the table article. If the information was not available, we searched among the studies cited by and citing the study of the table. Finally, we searched for information using keywords on the site web of science with (‘vernacular name’ OR ‘scientific name’) AND (mass OR weight OR height) AND (population name) for adult and by adding AND wean* for weaning. If the information was not available for the population we have left the box empty. For human studies, given that the study period is also considered, it is difficult to obtain information on mass and weight for the sample under consideration if the information is not directly in the article. Unfortunately, information on weight or height was not available in any of the human studies in the tables.